\documentclass[a4paper,11pt]{article}	
\pdfoutput=1 
\usepackage{jheppub} 
\usepackage{tikz}    
\usepackage{amsmath,amssymb,amsthm,amscd,graphicx,amsfonts}
\usepackage{autobreak,cancel}
\usepackage{enumerate}
\usepackage{ytableau}
\newcommand{\bitem}{\begin{itemize}}
\newcommand{\eitem}{\end{itemize}}
\newcommand{\be}{\begin{equation}}
\newcommand{\ee}{\end{equation}}
\newcommand{\ba}{\begin{aligned}}
\newcommand{\ea}{\end{aligned}}

\usepackage[table]{xcolor}

\usepackage{enumerate}
\usepackage{ytableau}

\usepackage{multirow}

\newcommand{\OO}{\mathbb O}
\newcommand{\CC}{\mathbb C}
\newcommand{\HH}{\mathbb H}

\newcommand{\RR}{\mathbb R}
\newcommand{\ZZ}{\mathbb Z}

\newcommand{\cN}{\mathcal N}
\newcommand{\cF}{\mathcal F}

\newcommand{\cT}{\mathcal T}

\newcommand{\Eh}{E_{7+1/2}}
\newcommand{\Dh}{D_{6+1/2}}
\newcommand{\Dhh}{D_{6+1/2+1/2}}

\newcommand{\mg}{\mathfrak{g}}

\newcommand{\Ehs}{E_{7\frac12}}
\newcommand{\Dhs}{D_{6\frac12}}
\newcommand{\Dhhs}{D_{6\frac12\!\frac12}}
\newcommand{\Ahs}{A_{5\frac12}}
\newcommand{\Chs}{C_{3\frac12}}
\newcommand{\ADs}{AD_{3\frac12}}
\newcommand{\AGs}{AG_{1\frac12}}
\newcommand{\UAs}{U\!A_{1\frac12}}

\title{\boldmath 2d Conformal Field Theories on Magic Triangle}
\author[a]{Kimyeong Lee}
\author[b]{and Kaiwen Sun}

\affiliation[a]{ 
Beijing Institute of Mathematical Sciences and Applications (BIMSA), 
Huaibei Town, Huairou District, Beijing 101408, China}
  
\affiliation[b]{School of Mathematical Sciences, University of Science and Technology of China, Hefei 230026, Anhui, China}

\emailAdd{klee@bimsa.cn}
\emailAdd{kwsun@ustc.edu.cn}

\abstract{
The magic triangle due to Cvitanovi\'c and Deligne--Gross is an extension of the Freudenthal--Tits magic square of semisimple Lie algebras. In this paper, we identify all two-dimensional rational conformal field theories associated to the magic triangle. These include various Wess--Zumino--Witten (WZW) models, Virasoro minimal models, compact bosons and their non-diagonal modular invariants. At level one, we  uncover a two-parameter family of fourth-order modular linear differential equation   whose solutions yield the affine characters of all elements in the magic triangle. We further establish  a universal coset relation for the whole triangle, generalizing   the dual-pair structure with respect to $(E_8)_1$ in the Cvitanovi\'c--Deligne exceptional series. This  coset structure determines  the dimensions and degeneracies  of all primary fields and leads to five atomic models  from which all theories in the triangle can be constructed.   At level two, we find that a distinghuished row of the triangle -- the subexceptional series-- exhibits emergent   $N=1$ supersymmetry. The corresponding   Neveu--Schwarz/Ramond characters satisfy a one-parameter family of fermionic modular linear differential equations. In addition, we find several new uniform coset constructions involving WZW models at higher levels.}

\begin{document}

\setcounter{tocdepth}{2}
\maketitle
\flushbottom

\section{Introduction}\label{sec:intro}
The magic triangle of groups was discovered in both theoretical physics \cite{Cvitanovic:1980ed} and mathematics \cite{DG}. It may be viewed as a triangular extension of the Freudenthal--Tits magic square $\mathfrak{m}(\mathbb{A}_1,\mathbb{A}_2)$ of semisimple Lie algebras and Lie groups, where $\mathbb{A}$ ranges over the four normed division algebras: the real numbers, complex numbers, quaternions, and octonions. From another perspective, the magic triangle can be regarded as a two-dimensional generalization of the celebrated Cvitanovi\'c--Deligne exceptional series \cite{Cvitanovic:1980ed,Cvitanovic:2008zz,Deligne}. 
The original magic triangle of Lie groups, found by Cvitanovi\'c \cite[Table 6]{Cvitanovic:1980ed} from certain Diophantine conditions on representation dimensions in the context of extended supergravity theories, is reproduced in Table~\ref{tb:MT0}. The magic triangle of Deligne and Gross is built as a system of centralizers from dual pairs in $E_8$, see \cite[Table 1]{DG}.

The magic triangle is symmetric, and each row (or, equivalently, each column) is referred to as an \emph{exceptional series}. In Table~\ref{tb:MT0}, the bottom row, terminating at $E_8$, is known as the \emph{Cvitanovi\'c--Deligne exceptional series}. The row immediately above it, ending at $E_7$, is called the \emph{subexceptional series}. The third row from the bottom, ending at $E_6$, is known as the \emph{Severi exceptional series}, while the fourth row from the bottom, terminating at $F_4$, is referred to as the \emph{Severi-section exceptional series}.

\begin{table}[ht]
\def\arraystretch{1.5}

	\centering
	\begin{tabular}{cccccccc|ccc}
       &  &   &  &   &   &     & $A_1$  & $1/3$ \\
    &  &   &  &   &   & $U_1$   & $A_2$  &  $1/2$\\  
 &   &  &   &  &     & $A_1$ &    $G_2$  & $2/3$\\  
 &   &   &  &   & $U_1^2$ & $A_1^3$ &    $D_4$ & $1$ \\  
    &   &  &     & \cellcolor{gray!20} $A_1 $ &\cellcolor{gray!20} $A_2$ &\cellcolor{gray!20} $C_3$ &     \cellcolor{gray!20}$F_4$ & $3/2$ & $\RR$ & Severi-section\\  
  &   &    & $U_1^2$  &\cellcolor{gray!20} $A_2$ & \cellcolor{gray!20}$A_2^2$   &\cellcolor{gray!20} $A_5$   & \cellcolor{gray!20}$E_6$  & $2$ & $\CC$ & Severi\\  
          & $U_1$ & $A_1$ & $A_1^3$ & \cellcolor{gray!20}$C_3$ & \cellcolor{gray!20}$A_5$ & \cellcolor{gray!20}$D_6$   &\cellcolor{gray!20} $E_7$  & $3$& $\HH$ & sub \\  
 $A_1$ & $A_2$  &  $G_2$   & $D_4$  & \cellcolor{gray!20} $F_4$          & \cellcolor{gray!20}$E_6$& \cellcolor{gray!20}$E_7$ & \cellcolor{gray!20}$E_8$ & $5$ & $\OO$ & Cvitanovi\'c--Deligne\!\!\\  \hline
   $1/3$ & $1/2 $& $2/3$ & $1$ & $3/2$ & $2$  & $3 $&   $5$ & $\nu\backslash \mu$  \\ 
 & &  & & $\RR$ & $\CC$ & $\HH$ & $\OO$  \\
		\end{tabular}
        \caption{The magic triangle of Lie groups of Cvitanovi\'c \cite{Cvitanovic:1980ed} and Deligne--Gross \cite{DG}. The lower right $4\times 4$ square is the Freudenthal--Tits magic square, where $a=2(\nu-1)$ is the dimension of the division algebra. The last four rows from the bottom up are Cvitanovi\'c--Deligne, sub, Severi and Severi-section exceptional series respectively.}
			\label{tb:MT0}
		\end{table}

Each exceptional series exhibits distinctive and remarkable properties in representation theory. For instance, the Cvitanovi\'c--Deligne exceptional series displays striking uniformity in the tensor product decompositions of adjoint representations, and the dimensions of its irreducible representations admit expressions as rational functions of the dual Coxeter number \cite{Deligne,Cohen,deligne1996serie}. This series appears frequently in theoretical physics, including 2d rational CFTs \cite{Mathur:1988na}, 4d $N=2$ superconformal field theories (SCFTs) (see e.g. \cite[Section 4]{Beem:2017ooy}),  and 6d $N=(1,0)$ SCFTs \cite{Morrison:2012np}. The subexceptional series first appeared as global symmetries in 4d extended supergravity theories \cite{Cvitanovic:1980ed}, and it also possesses many distinguished representations with uniform tensor product decompositions and dimension formulas \cite{LMseries}. From the perspective of Vogel’s theory of the universal Lie algebra, both the Cvitanovi\'c--Deligne series and the subexceptional series form straight lines in Vogel’s projective plane \cite{vogel1999universal}.
Moreover, the four exceptional series are deeply connected with geometry \cite{LM02,LMseries,LANDSBERG2001477}. In particular, the terminology Severi exceptional series and Severi-section exceptional series reflects their geometric relationship with Severi varieties \cite{LMseries}.

Several conceptual issues arise in the original magic triangle displayed in Table \ref{tb:MT0}. First, despite its name, the original magic triangle of Lie groups does not form a complete triangular array, as certain entries along the slope are missing. These absent elements were identified as discrete groups in \cite{DG}. A second issue concerns the common practice in the physics literature of adjoining an additional entry ``Lee--Yang model'' at the beginning of the Cvitanovi\'c--Deligne exceptional series \cite{Mathur:1988na,Beem:2017ooy}, which further obscures the triangular structure. A third question is whether the famous intermediate Lie algebra $\Eh$ \cite{LM}, lying between $E_7$ and $E_8$, can be incorporated into the Cvitanovi\'c--Deligne exceptional series in a manner compatible with the magic triangle. In this work, we address these three issues in a unified framework and show that their resolution naturally requires an extension of the original diagram, resulting in a genuinely triangular structure.

Our approach is based on two-dimensional rational conformal field theories (RCFTs) associated with the magic triangle. This may be viewed as a natural generalization of the work of Mathur, Mukhi, and Sen on the  2d RCFTs associated with the Cvitanovi\'c--Deligne exceptional series \cite{Mathur:1988na,Mathur:1988gt}. From this perspective, the resulting structure can be interpreted as an affine analogue of the original magic triangle. The construction is particularly natural at level one; the resulting extended triangle is displayed in Table~\ref{tb:MT}. For each Lie group appearing in the original magic triangle in Table~\ref{tb:MT0}, the corresponding entry in Table~\ref{tb:MT} is given by the associated Wess--Zumino--Witten model mostly at level one. The remaining entries consist of various minimal models, compact boson theories, and their non-diagonal modular invariants, which will be explained in detail below. In particular, we introduce an additional row ending at $E_{7+1/2 }$, referred to as the \emph{intermediate exceptional series}, which has been studied systematically in \cite{Lee:2024fxa}.

\begin{table}[ht]
\def\arraystretch{2}

	\centering
	\resizebox{\linewidth}{!}{
	\begin{tabular}{c cccccccccc|c}
  & &   &  &   &  &   &   &  &    &  0 & 1/5 \\     
 & &   &  &   &  &   &   &  &   0 & $\mathrm{Vir}_{5,2}^{\rm eff}$ &  1/4  \\  
&  &   &  &   &  &   &   & 0 &  $\mathrm{Vir}_{5,3}^{\rm eff}$  & $(A_1)_1$ &  1/3  \\  
& &   &  &   &  &   &  0 & $(U_1)_3$ &  $(\UAs)_1$ & $(A_2)_1$ & 1/2 \\  
& &   &  &   &  &  0 & $\mathrm{Vir}_{6,5}^{\rm e}$ & $(A_1)_3$ &  $(\AGs)_1$ & $(G_2)_1$ &  2/3  \\  
 &   &   &  &   & 0 & $D_{2\rm A}$    & $M_2$ & $(A_1^{\oplus 3})_1$ &  $(\ADs)_1$ & $(D_4)_1$ & 1\\   &   &   &  & 0  &  $D_{2\rm A}$ & $(A_1)_8^{\rm e}$ & $(A_2)_2$ & $(C_3)_1$ &   $(\Chs)_1$ & $(F_4)_1$ & 3/2\\  
 & &   &  0   & $\mathrm{Vir}_{6,5}^{\rm e}$  &  $M_2 $  & $(A_2)_2$ & $(A_2^{\oplus 2})_1$   & $(A_5)_1$  & $(\Ahs)_1$ & $(E_6)_1$ & 2 \\  
 &    &   0  & $(U_1)_3$ & $(A_1)_3$ & $(A_1^{\oplus 3})_1$ & $(C_3)_1$ & $(A_5)_1$ & $(D_6)_1$ & $(\Dhs)_1$ & $(E_7)_1$ & 3 \\  
 &    0  &  $\mathrm{Vir}_{5,3}^{\rm eff}$   & $(\UAs)_1$ & $(\AGs)_1$ & $(\ADs)_1$ & $(\Chs)_1$ & $(\Ahs)_1$ & $(\Dhs)_1$ & $(\Dhhs)_1$ & $(\Ehs)_1$ & $4$\\  
0 & $\mathrm{Vir}_{5,2}^{\rm eff}$ & $(A_1)_1$ & $(A_2)_1$  &  $(G_2)_1$   & $(D_4)_1$  &  $(F_4)_1$          & $(E_6)_1$& $(E_7)_1$&$(\Ehs)_1$& $(E_8)_1$ & 5\\ \hline
$1/5$ &  $1/4$ & $1/3$ & $1/2 $& $2/3$ & $1$ & $3/2$ & 2  & 3 & 4 & 5 & $\nu\backslash \mu$  \\  
 \end{tabular}}
        \caption{2d CFTs associated to the extended magic triangle at level one.}
			\label{tb:MT}
		\end{table}

The main results of this work can be summarized as follows.

\emph{1. We determine the 2d RCFT associated with each entry of the extended magic triangle at level one, as  presented in Table \ref{tb:MT}. The characters of these RCFTs satisfy a uniform two-parameter family of fourth-order holomorphic modular linear differential equations (MLDEs), given in \eqref{MLDE1}.}

\emph{2. The magic coset \eqref{magiccoset} provides a natural generalization of the dual-pair structure with respect to $(E_8)_1$ in the Cvitanovi\'c--Deligne exceptional series to the entire magic triangle.}

\emph{3. The dimensions, degeneracies, and modular $S$-matrices of all theories in the magic triangle at level one are given in a  uniform manner.}

\emph{4. We show that all thirty theories on the level-one magic triangle can be decomposed into five elementary building blocks: 
$$
{\rm Vir}^{\rm eff}_{5,2},\ {\rm Vir}^{\rm eff}_{5,3},\  (U_1)_3,\  {\rm Vir}^{\rm e}_{6,5},\  D_{\rm 2A}.
$$} 

\emph{5. The subexceptional series at level two exhibits ${N}=1$ supersymmetry. The corresponding fermionic characters satisfy a one-parameter family of fourth-order fermionic MLDEs.}

\emph{6. We identify many uniform coset constructions involving exceptional series at level two such as $(\mathfrak{g}_{\rm CD})_2/{(\mathfrak{g}_{\rm sub})_2\times (A_1)_2}$ and $(\mathfrak{g}_{\rm sub})_2/{(\mathfrak{g}_{\rm Severi})_2\times (U_1)_6}$.}

This paper is organized as follows. In Section~\ref{sec:2}, we review background material on exceptional series, the magic square, and the magic triangle, and provide a brief introduction to characters of 2d RCFTs and modular linear differential equations (MLDEs). In particular, we recall the seminal result of Mathur, Mukhi, and Sen \cite{Mathur:1988na} that the CFT-type solutions of second-order holomorphic MLDEs are precisely the level-one affine characters of the Cvitanovi\'c--Deligne exceptional series.
In Section~\ref{sec:3}, we extend the framework of Mathur--Mukhi--Sen to the full magic triangle and investigate several structural properties of this generalization, which may be regarded as an affine realization of the magic triangle at level one. Section~\ref{sec:4} is devoted to the magic triangle at level two, with particular emphasis on the exceptional series appearing in the lower rows which exhibits richer symmetry structures.  In Section~\ref{sec:5}, we discuss the magic triangle at arbitrary positive integer levels.
In the end, we give some concluding remarks. In Appendix \ref{sec:A}, we explicitly present the data for the extended magic triangle and the CFTs at level one. In Appendix \ref{sec:B0}, we explicitly give the $S$-matrices for the CFTs at level one of Cvitanovi\'c--Deligne exceptional series. In Appendix \ref{sec:B}, we discuss the spurious solutions of the 4th order MLDE for Cvitanovi\'c--Deligne exceptional series.

We use $()^{\rm e}$ for a non-diagonal modular invariant of RCFT, that is,  a VOA extension. For example, ${\rm Vir}^{\rm e}_{6,5}$ is a non-diagonal modular invariant of Virasoro minimal model ${\rm Vir}_{6,5}$ that describes the critical three-state Potts model. We denote the RCFT associated with the triangle element $(\mu,\nu)$ at level $k$ as $\mathcal{T}^{(\mu,\nu)}_{k}$. In particular, we denote the level-one theory as $\mathcal{T}^{(\mu,\nu)} $ for simplicity. We use  subscript numbers to denote the degeneracies of primary fields, and we use the superscript ``$\rm eff$" to denote the effective theory of non-unitary minimal models.

\section{Review of the magic triangle}\label{sec:2}
\subsection{Exceptional series and magic triangle}

We begin by introducing a key parameter $\nu$ associated with the magic triangle and the exceptional series. Throughout this paper, $\nu$ takes values in the finite set
\begin{align}\label{eq:vlist}
\{1/5,1/4,1/3,1/2,2/3,1,3/2,2,3,4,5\}.
\end{align} 
These values  correspond to the following extended Cvitanovi\'c--Deligne exceptional series, respectively:
\begin{align}\label{eCD}
e\subset A_{1/2}\subset  A_1\subset A_2 \subset G_2 \subset D_4 \subset F_4 \subset E_6 \subset E_7 \subset \Eh \subset E_8. 
\end{align} 
Here $e$  denotes a trivial element. For the simple Lie algebras $\mathfrak{g}$ in the series, which constitute the original Cvitanovi\'c--Deligne exceptional series, the dimension and the dual Coxeter number \cite{Cvitanovic:1980ed,Cvitanovic:2008zz,Deligne} are expressed in terms of  $\nu$ as   \begin{align}\label{eq:CDdim}
    d_{\mg}&=\frac{2(5\nu-1)(6\nu+1)}{(\nu+1)},\\   \label{eq:CDh} h^\vee_{\mg} &=6\nu.
\end{align}
Moreover, infinitely many representations of $\mg$ admit dimension formulas that can be expressed as rational functions of $\nu$ \cite{Cohen,LM}. The elements in \eqref{eCD} form several dual pairs with respect to $E_8$: $(e,E_8)$, $(A_{1/2},\Eh)$, $(A_1,E_7)$, $(A_2,E_6)$, $(G_2,F_4)$ and $(D_4,D_4)$. In each pair, the $\nu$ parameters of the two elements are inverse of each other.

The original Cvitanovi\'c--Deligne exceptional series can be  extended to include intermediate Lie algebras $A_{1/2}$ \cite{ShtepinAC} and $E_{7+1/2}$ \cite{LM}, whose dual Coxeter number and dimension obey the same formulas \eqref{eq:CDdim} and \eqref{eq:CDh}.\footnote{The notation “half” indicates that the dual Coxeter number of the intermediate Lie algebra is the arithmetic mean of those of the two algebras appearing in the corresponding inclusions, namely $A_0\subset A_{1/2}\subset A_1$ and $E_7\subset \Eh\subset E_8$.} To be specific, $A_{1/2}$ has dual Coxeter number $3/2$ and dimension $1$. In many senses, one may regard $A_{1/2}$  as Lie superalgebra $osp(1|2)$ which has dual Coxeter number $3/2$ and superdimension $1$. Then affine characters of $(A_{1/2})_k$ are just the supercharacters of $osp(1|2)_k$. An alternative interpretation is to regard $A_{1/2}$ as a formal object, denoted $LY$, such that the associated theory $(LY)_k$ realizes the Lee--Yang model at level $k$, namely the effective Virasoro minimal model $\mathrm{Vir}^{\rm eff}_{(2k+3,2)}$ \cite{Duan:2022ltz}. This correspondence follows from the fact that the supercharacters of $\mathfrak{osp}(1|2)_k$ agree, as $q$-series, with the characters of $\mathrm{Vir}^{\rm eff}_{(2k+3,2)}$ \cite{Ferrari:2023fez}. On the other hand, the $\Eh$ is a famous intermediate Lie algebra with dual Coxeter number $24$ and dimension $190$. Its existence was noted both in physics \cite{Mathur:1988na,Cvitanovic:2008zz} and mathematics \cite{Cohen}, and it was subsequently constructed by Landsberg and Manivel using sextonions \cite{LM}. Many irreducible representations of $\Eh$ have been studied in \cite{Lee:2023owa}. The affine characters of $(\Eh)_1$ were found both by MLDE \cite{Mathur:1988na} and Hecke operator \cite{Harvey:2018rdc}. The affine characters of $(\Eh)_2$ were recently found by Hecke operator in \cite{Duan:2022ltz}.  

The magic triangle is a two-dimensional symmetric generalization of the Cvitanovi\'c--Deligne exceptional series, parameterized by a pair $(\mu,\nu)$ with $\mu\nu\ge 1$. In the case $\mu\nu = 1$, all entries reduce to trivial elements due to the zero dimension.\footnote{They may nevertheless be interesting, and we hope to return to this point elsewhere.} One systematic construction of the magic triangle proceeds by placing the Cvitanovi\'c--Deligne exceptional series along the bottom row and the rightmost column, and then defining each remaining entry as the centralizer of the dual of corresponding column element inside the row element \cite{DG}.   
The original magic triangle introduced in \cite[Table~6]{Cvitanovic:1980ed} is reproduced in Table~\ref{tb:MT0}. The subarray with $\mu,\nu>1$, highlighted in Table~\ref{tb:MT0}, is known as the Freudenthal--Tits magic square. The Lie algebras appearing in the magic square admit a uniform construction in terms of pairs of normed division algebras, denoted by $\mathfrak{m}(\mathbb{A}_1,\mathbb{A}_2)$ for $\mathbb{A}=\mathbb{R},\mathbb{C},\mathbb{H},\mathbb{O}$. 
This construction realizes the idea that all five exceptional Lie algebras are related to the octonions; see \cite{Baez:2001dm} for a comprehensive review. 
Upon including the sextonion algebra $\mathbb{S}$, the magic square can be further extended to incorporate several intermediate Lie algebras; see \cite{Westbury:2004vav,Lee:2024fxa} for a detailed discussion. In this context, it is sometimes convenient to replace the parameter $\nu$ by $a=2(\nu-1)$, which takes values in the set $ 
    \{ -8/5,-3/2,-4/3,-1,-2/3,0,1,2,4,6,8\} .$
Here the numbers $1, 2, 4, 6$, and $ 8$ are the real dimensions of $\mathbb{R},\mathbb{C},\mathbb{H},\mathbb{S}$, and $\mathbb{O}$, respectively.  
Finally, it is worth noting that the magic square has also appeared in the study of supergravity theories, see for example \cite{Anastasiou:2013hba}. 

For the element ${\mathfrak{g}}^{(\mu,\nu)}$ in the original magic triangle in Table \ref{tb:MT0}, the dimension and dual Coxeter number admit the following universal formulas:
\begin{align}\label{eq:dim}
    d_{\mathfrak{g}}^{(\mu,\nu)}&= \frac{12(\mu\nu-1)(\mu\nu+\mu+\nu-4)}{(\mu+1)(\nu+1)} , \\ \label{eq:dualCoxeter}
    h^\vee_{\mathfrak{g}}{}^{(\mu,\nu)}&=  \mu\nu+ \mu+ \nu -5. 
\end{align}
The dimension formula was given in \cite[Equation (11.1)]{Cvitanovic:1980ed} in a different form, and also given for magic square in \cite[Proposition 3.2]{LM}. In particular, by specializing to $\mu = 5$, one recovers the expressions for the dimensions \eqref{eq:CDdim} and dual Coxeter numbers \eqref{eq:CDh} of the Cvitanovi\'c--Deligne exceptional series. 
The original magic triangle of Cvitanovi\'c and Deligne--Gross does not include the cases $\mu,\nu = 1/4$ or $4$. Incorporating these values naturally leads to an extended magic triangle. Interestingly, the universal formulas \eqref{eq:dim} and \eqref{eq:dualCoxeter} continue to hold throughout this extension. In particular, the dimensions are all non-negative integers. 
For $\nu = 1/4$, the extension introduces a single additional element, namely $A_{1/2}$. For $\nu = 4$, one obtains the intermediate exceptional series, whose dimensions and dual Coxeter numbers were determined in \cite{Lee:2024fxa} and are consistent with the general formulas \eqref{eq:dim} and \eqref{eq:dualCoxeter}. The dual Coxeter numbers and dimensions of all entries in the extended magic triangle are summarized in Table~\ref{tb:MTdata0}.

Several entries of the extended magic triangle require additional clarification. The formal groups ${\mathfrak{g}}^{(3/2,1)}$ and ${\mathfrak{g}}^{(2,2/3)}$ both have vanishing dimension and dual Coxeter number equal to $-1$. In \cite{DG}, these entries are realized concretely by the finite groups $\mathbb{Z}_2 \times \mathbb{Z}_2$ and $\mathbb{Z}_3$, respectively. 
The element ${\mathfrak{g}}^{(4,1/3)}$ has dimension $1$ and dual Coxeter number $2/3$. It may be interpreted as a formal object, denoted $IM$, such that the associated theory $(IM)_k$ realizes the effective minimal model $\mathrm{Vir}^{\rm eff}_{(3k+2,3)}$ \cite{Lee:2024fxa}. There also exist three entries whose underlying Lie algebras have standard dimensions but nonstandard dual Coxeter numbers. These are  ${\mathfrak{g}}^{(3,2/3)}=A_1$, ${\mathfrak{g}}^{(3/2,3/2)}=A_1$ and ${\mathfrak{g}}^{(2,3/2)}=A_2$. According to the general formula \eqref{eq:dualCoxeter}, their dual Coxeter numbers are given by $2/3$, $1/4$, and $3/2$, respectively. This means that the invariant bilinear forms on these Lie algebras are rescaled by factors of $3$, $8$, and $2$. Consequently, when constructing the associated Wess--Zumino--Witten models, the levels must be rescaled by the same factors, namely $3$, $8$, and $2$, respectively.

\subsection{Modular linear differential equation}
In this section, we briefly review modular linear differential equations (MLDEs) and their fermionic counter parts, namely MLDEs associated with congruence subgroups of level two. There exist several equivalent formulations of MLDEs, including those based on ordinary derivatives, Serre derivatives, and Rankin--Cohen brackets; see, for example, \cite{nagatomo2022modular}. For our purposes, it is most convenient to work with the Serre derivative
\begin{equation}
    D_k = q\frac{d}{d q} - \frac{k}{12}E_2,
\end{equation}
which maps a holomorphic modular form of weight $k$ to a holomorphic modular form of weight $k+2$. A modular linear differential equation of degree $d$ can be written in the form
\begin{equation}\label{eq:mlde}
    [\phi_{2l}D^d+\phi_{2l+2}D^{d-1}+\phi_{2l+4}D^{d-2}+\dots+\phi_{2l+2d}]\chi=0,
\end{equation}
where the coefficients $\phi_k$ are holomorphic modular forms of weight $k$ for $SL(2,\mathbb{Z})$, generated by the Eisenstein series $E_4$ and $E_6$. The non-negative integer $l$ is referred to as the \emph{index} of the MLDE. When $l=0$, the equation is called holomorphic or monic; otherwise meromorphic.

We are interested in the special situation where $\chi$ is a modular form of weight zero on certain congruence subgroup of $SL(2,\mathbb{Z})$.  
In particular, we focus on solutions of the form 
$$
q^{\alpha}(a_0+a_1q+a_2q^2+\dots),
$$
where $\alpha$ is a rational number, and all Fourier coefficients $a_i$ are non-negative integers. Such solutions are referred to as \emph{CFT-type} solutions of MLDEs and constitute a necessary condition for $\chi$ to be admissible as a character of a two-dimensional rational conformal field theory (RCFT). 
The modular properties of RCFT characters play a vital role in the study of two-dimensional conformal field theories. The full set of characters of a 2d RCFT forms a vector-valued modular form for $SL(2,\mathbb{Z})$ \cite{Zhu}, and the congruence subgroup property of RCFT characters was established in \cite{Ng:2012ty}. The systematic study of MLDEs satisfied by RCFT characters was initiated by Eguchi and Ooguri \cite{Eguchi:1987qd} and further developed in \cite{Anderson:1987ge,Mathur:1988na}.

Suppose that the MLDE \eqref{eq:mlde} admits $d$ linearly independent CFT-type solutions forming a vector-valued modular form. The Wronskian analysis of \cite{Mathur:1988na} implies that the index $l$ is related to the degree $d$ and the exponents $\alpha_i$ by
\begin{equation}
    \label{eq:index}
\frac{l}{6} = \frac{d(d-1)}{12} - \sum_{i = 0}^{d-1} \alpha_i.
\end{equation}
When these solutions are identified with the characters of a 2d RCFT, the exponents are related to the central charge $c$ and conformal weights $h_i$ of primary fields by
$$
\alpha_i=-\frac{c}{24}+h_i.
$$
One of the important applications of MLDEs is the classification of 2d RCFTs with a given number of characters, as well as the bootstrapping of new 2d RCFTs. This program requires a classification of CFT-type   solutions of MLDEs for given pairs $(d,l)$. In the next subsection, we review the classical work of Mathur, Mukhi, and Sen on MLDEs with $(d,l) = (2,0)$ \cite{Mathur:1988na}. More recent progress on the classification of CFT-type MLDE solutions can be found, for example, in \cite{Kaidi:2021ent}.

It is natural to generalize the properties of the characters of 2d RCFTs to the case of 2d fermionic RCFTs. In the presence of fermions, one must distinguish between periodic (Ramond) and antiperiodic (Neveu--Schwarz) boundary conditions along a  circle, leading to four sectors depending on the boundary conditions along the two cycles of the torus. In summary, fermionic RCFTs admit four types of characters: NS, $\widetilde{\rm NS}$, R, and $\widetilde{\rm R}$. They transform as vector-valued modular forms for the congruence subgroups $\Gamma_\theta(2)$, $\Gamma^0(2)$, $\Gamma_0(2)$, and $SL(2,\mathbb{Z})$, respectively; see, for instance, 
\cite{Bae:2020xzl,Duan:2022kxr} for details. The fermionic characters of a fermionic RCFT satisfy corresponding fermionic MLDEs. For example, the NS characters obey an MLDE of the form \eqref{eq:mlde}, where the coefficients $\phi_k$ are holomorphic modular forms of weight $k$ for the congruence subgroup $\Gamma_\theta(2)$. Then the fermionic MLDEs for $\widetilde{\rm NS}$ characters and the one for R characters can be obtained by modular transformation from the one for NS characters.

\subsection{Mathur--Mukhi--Sen's discovery}\label{sec:MMS}
The simplest nontrivial holomorphic modular linear differential equations are of second order. In \cite{Mathur:1988na,Mathur:1988gt}, Mathur, Mukhi, and Sen classified all CFT-type solutions of the second-order holomorphic MLDE
\begin{equation}\label{KZ}
    [D^2+\lambda E_4]\chi=0.
\end{equation}
This equation is also known in the literature as the Kaneko--Zagier equation \cite{KZ}. Mathur, Mukhi, and Sen showed that   only for ten discrete values of the parameter $\lambda$ does the MLDE \eqref{KZ} admit CFT-type solutions satisfying with the unique vacuum condition. Moreover, for each such value of $\lambda$, the corresponding solutions coincide precisely with the characters of level-one Wess--Zumino--Witten models associated with the extended Cvitanovi\'c--Deligne exceptional series: 
\begin{align}\label{MMS}
    \mathrm{Vir}_{5,2}^{\rm eff},\ (A_1)_1,\ (A_2)_1,\ (G_2)_1,\  (D_4)_1,\  (F_4)_1,\ (E_6)_1,\   (E_7)_1,\   (\Eh)_1,\  (E_8)_1.
\end{align}
We refer to the sequence \eqref{MMS} as the \emph{Mathur--Mukhi--Sen series}, and denote these theories as $\cT^{(\nu,5)}$ or $\cT^{(5,\nu)}$, where  $\nu$ is defined earlier and is associated with the CD exceptional series. This notation will be consistent with magic triangle at level one in Table \ref{tb:MT}. The associated CFT data are summarized in Table~\ref{tb:CD}. Here $\mathrm{Vir}_{5,2}^{\rm eff}$ denotes the effective Lee--Yang model with effective central charge $2/5$.\footnote{This is to exchange the labeling of the vacuum and non-vacuum operator of the original Lee--Yang model with $c=-{22}/{5}.$} 
The entry $(\Eh)_1$ corresponds to the solution of the MLDE with central charge $c = 38/5$. Mathur, Mukhi, and Sen observed that, if this solution were interpreted as a level-one WZW model, the corresponding symmetry algebra would have dimension $190$ and dual Coxeter number $24$, and hence could not arise from a semisimple Lie algebra. It was subsequently shown that these solutions can be realized as the characters of an intermediate vertex subalgebra of $(E_8)_1$ \cite{Kawasetsu}. Accordingly, $(\Eh)_1$ is often reffered to as an intermediate VOA in the physics literature. An alternative interpretation identifies the solutions as the characters of Ramond-twisted modules of a minimal $\mathcal{W}$-algebra of type $E_8$ with central charge $32/5$ \cite{Kawasetsu:2018irs}. Finally, the $(E_8)_1$ theory has the single character, giveb by $j(\tau)^{1/3}$ \cite{Mathur:1988na}. In this case, the second order MLDE admits a spurious solution, see Appendix \ref{sec:B}.

\begin{table}[ht]
\def\arraystretch{1.5}

	\centering
	
	\begin{tabular}{c| cccccccccc}

$\nu$ & $1/4$ & $1/3$ & $1/2 $& $2/3$ & $1$ & $3/2$ & 2  & 3 & 4 & 5   \\  \hline
 & $\mathrm{Vir}_{5,2}^{\rm eff}$ & $(A_1)_1$ & $(A_2)_1$  &  $(G_2)_1$   & $(D_4)_1$  &  $(F_4)_1$          & $(E_6)_1$& $(E_7)_1$&$(\Ehs)_1$& $(E_8)_1$ \\ 
$c$ & $\frac25$ & 1 & 2  & $\frac{14}{5}$ & 4 & $\frac{26}{5}$  & 6 & 7 & $\frac{38}{5}$ & 8  \\
$h$ & $\frac15$ & $\frac14$ & $\frac13$  & $\frac25$ & $\frac12$ & $\frac35$  & $\frac23$ & $\frac34$ & $\frac45$ & $- $  \\
$\cN$  & 1 & 1 & 2  &1  & 3 & 1  &2  &1  &1  & $-$ \\
$R$ & 1 & 2 & 3& 7 & 8 & 26 & 27 & 56& 57 & $-$ \\
 \end{tabular}
        \caption{2d CFTs associated to the extended Cvitanovi\'c--Deligne exceptional series at level $1$.}
			\label{tb:CD}
		\end{table}

Recall that the Wess--Zumino--Witten model $(\mathfrak{g})_k$ has central charge and conformal weights given by
\begin{equation}\label{eq:WZWc}
    c=\frac{k d_{\mathfrak{g}}}{k+h^\vee},\qquad h_\lambda=\frac{C_2(R_\lambda)}{2(k+h^\vee)},
\end{equation}
where $\lambda$ is the weight and $C_2(R_\lambda)$ is the quadratic Casimir invariant of representation $R_\lambda$. For the extended Cvitanovi\'c--Deligne exceptional series, both the dimension $d_{\mathfrak{g}}$ and the dual Coxeter number $h^\vee$ can be expressed as functions of the parameter $\nu$; the same is true for  the quadratic Casimir invariant of the fundamental representation. At level one, the corresponding central charge and the conformal weight of the non-vacuum primary field are given by
\begin{equation}
    c^{(\nu)}=\frac{2(5\nu -1)}{(\nu+1)},\qquad  h^{(\nu)}=\frac{\nu}{\nu+1}.
\end{equation}
In terms of the parameter $\nu$, the associated second-order holomorphic MLDE can be written compactly as $\Delta_\nu \chi = 0$, where
\begin{align}\label{eq:deltanu}
\Delta_{\nu} =      D^2-\frac{(5 \nu-1) (7 \nu+1)}{144 (\nu+1)^2}E_4 .
\end{align}
The modular invariant partition function of the 2d RCFT on torus takes the form
\begin{align} Z= |\chi_0|^2+{\mathcal N}|\chi_1|^2, \label{eq:CDpart} \end{align}
where $\mathcal N$ is a natural number, capturing the degeneracy of the non-vacuum primary operator. The characters of the Mathur--Mukhi--Sen series admit the Fourier expansions
\begin{align}\label{eq:CDchi}
    \chi_0 &= q^{-\frac{c}{24}}(1+ d_\mathfrak{g} q+ \cdots), \nonumber \\
    \chi_1 &= q^{-\frac{c}{24}+h}(R  + \cdots ),
\end{align}
where $R$ denotes the dimension of the fundamental representation of $\mathfrak{g}$. The degeneracy $\cN$ and dimension $R$ for various $\nu$ are collected in Table \ref{tb:CD}. 
The $S$-matrix arises from the modular transformation of the characters,
\begin{align}
    \left(\begin{array}{c} \chi_0 \\ \chi_1 \end{array}\right) (-1/\tau) = S\cdot \left(\begin{array}{c} \chi_0 \\ \chi_1 \end{array}\right) (\tau).
\end{align}
As a $2\times 2$ matrix not necessarily symmetric, the $S$-matrix satisfies 
$  \det S_{(\nu)}=-1$ and 
\begin{align} 
& S_{(\nu)}^T \cdot  \left(\begin{array}{cc}
    1 & 0 \\
    0 &  \cN_{(\nu)}
\end{array}\right)\cdot S_{(\nu)} =  \left(\begin{array}{cc}
    1 & 0 \\
    0 &  \cN_{(\nu)}
\end{array}\right). \label{eq:SId}\end{align}
The explicit form of the $S$-matrices for the Mathur--Mukhi--Sen series is presented in Appendix~\ref{sec:B0}. More generally, one may consider the full modular $S_{(\nu)}$-matrix as a $(1+\mathcal{N}_{(\nu)})$-dimensional unitary matrix.

It is well known that for the CD exceptional series at level $1$, the RCFTs pair together as a coset with respect to $\mathcal{T}^{(5,5)}=(E_8)_1$ \cite{Mathur:1988na}.  These coset relations are 
\begin{align}
    (E_7)_1=\frac{(E_8)_1}{(A_1)_1},\quad 
    (E_6)_1=\frac{(E_8)_1}{(A_2)_1},\quad 
    (F_4)_1=\frac{(E_8)_1}{(G_2)_1},\quad 
    (D_4)_1=\frac{(E_8)_1}{(D_4)_1},\quad 
\end{align}
In our notation, these cosets can be written uniformly as
\begin{align}\label{E8evel1}
    \mathcal{T}^{(5, \nu)}=\frac{ \mathcal{T}^{(5,5)}}{\mathcal{T}^{(1/\nu,5)}}.
\end{align}
The central charge and non-vacuum conformal weight $h_{\nu}$ of $\mathcal{T}_{5,\nu}$ theory satisfy the relations
\begin{align}
    c^{(\nu)}+c^{(1/\nu)}&=8, \nonumber \\
    h^{(\nu)}+h^{(1/\nu)}&=  1.
\end{align}
The degeneracy $\cN_{(\nu)}$ and $S$-matrix $S_{(\nu)}$ satisfy
\be \cN_{(\nu)}= \cN_{(1/\nu)}, \qquad  S_{(\nu)}= S_{(1/\nu)}.\label{eq:NSId} \ee
The characters satisfy the relation%
\begin{align}
    \chi_0^{(\nu)}\chi_0^{(1/\nu)} + \cN_{(\nu)}\chi_1^{(\nu)} \chi_1^{(1/\nu)} = j^{1/3}. \label{eq:CDcoset}
\end{align}

%
In addition, the coefficients in    the subleading $q$-order implies  
\begin{equation} \label{eq:bilinear} \cN_{(\nu)} R_{(\nu)} R_{(1/\nu)} = 312-60(\nu+1/\nu). 
\end{equation}
From the study of the solution in terms of the hypergeometric function \cite{Naculich:1988xv} with $h=\nu/(\nu+1)$ being the conformal weight of the non-vacuum primary operator, the explicit formula for $\cN_{(\nu)} R_{(\nu)}^2$ is known as 
\begin{align}
  \cN_{(\nu)}R_{(\nu)}^2 = 2^{16h} (4\sin^2(\pi h)-1) \bigg[ \frac{\Gamma(1-2h)\Gamma(1/2+3h)}{\Gamma(1+2h)\Gamma(1/2-h)}\bigg]^2. 
\end{align}
By multiplying $\cN_{(\nu)} R_{(1/\nu)}^2$ one can also obtain the relation \eqref{eq:bilinear}.

The characters of WZW models for the CD exceptional series at level one possess many other interesting properties such as Hecke relations \cite{Harvey:2018rdc}. 
Besides, it is recently shown that the vacuum character of WZW models for CD exceptional series at level one can be constructed from the monodromy trace of some Argyres--Douglas theories \cite{Kim:2024dxu}.

\section{Magic triangle at level one}\label{sec:3}

A principal motivation of this work is to extend the discovery of Mathur, Mukhi, and Sen concerning second-order holomorphic MLDEs and the Cvitanovi\'c--Deligne exceptional series to the full magic triangle. To this end, we begin by analyzing the level-one Wess--Zumino--Witten models associated with the ordinary Lie groups appearing in the original magic triangle of Table~\ref{tb:MT0}. For each resulting rational conformal field theory, we compute the characters and find that they satisfy a uniform two-parameter family of fourth-order holomorphic MLDE \eqref{MLDE1}. 
Assuming that this MLDE framework extends to the enlarged magic triangle, we determine the characters corresponding to the previously unknown entries and identify the associated 2d RCFTs. The complete set of resulting theories is summarized in Table~\ref{tb:MT}, which we interpret as the extended magic triangle at level one. We further uncover several elegant properties of this level-one triangle, including the magic coset structure,   matrix-valued modularity of the characters with explicit S-matrices, the dimensions and degeneracies of   primary operators,
and the existence of five atomic models that together  constitute all theories in the triangle.

\subsection{The fourth order MLDE}
Let us first consider the level-one theories for the original magic triangle in Table \ref{tb:MT0}. 
Substituting the dimension and dual Coxeter number formulas \eqref{eq:dim}, \eqref{eq:dualCoxeter} into the central charge formula of WZW model \eqref{eq:WZWc}, we obtain the central charge of the level-one theory $\mathcal{T}^{(\mu,\nu)}$ as
\begin{align}
    c^{(\mu,\nu)}=  \frac{12(\mu\nu -1)}{(\mu+1)(\nu+1)}. \label{eq:ccharge}
\end{align}
We observe that each theory $\mathcal{T}^{(\mu,\nu)}$ admits at most four distinct characters. With the exception of the Cvitanovi\'c--Deligne exceptional series, the theory $\mathcal{T}^{(\mu,\nu)}$ with $\mu\neq\nu$ generically possesses four inequivalent characters. Their conformal weights take the universal form
\begin{align}
    h_i^{(\mu,\nu)}=0, \ \frac{\mu\nu-1}{(\mu+1)(\nu+1)} , \  \frac{\nu}{\nu+1},\  \frac{\mu}{\mu+1}.
    \label{eq:weight}
\end{align}

The exponents $\alpha_i=-c/24+h_i$, appearing in the corresponding characters $\chi_i$, take values 
\begin{align}\label{eq:alphai}
    \alpha_0=-\alpha_1=\frac{1-\mu\nu}{2(\mu+1)(\nu+1)},\quad \alpha_2=\frac{\mu\nu+2\nu+1}{2(\mu+1)(\nu+1)}, \quad \alpha_3=\frac{\mu\nu+2\mu+1}{2(\mu+1)(\nu+1)}.
\end{align}   
We denote the four characters of the theory $\mathcal{T}^{(\mu,\nu)}$ by
\begin{align}
 \chi_0&=  q^{\alpha_0}    \Big( 1 +\sum_{n=1}^\infty a_n q^n \Big)    \ \   {\rm with} \ a_1=d_\mathfrak{g} ,   \\
\chi_1&=  q^{\alpha_1}       \Big( R_1 +\sum_{n=1}^\infty b_n q^n \Big) ,   \\
\chi_2&=  q^{\alpha_2}      \Big( R_2 +\sum_{n=1}^\infty c_n q^n \Big) ,   \\
\chi_3&=  q^{\alpha_3}     \Big( R_3 +\sum_{n=1}^\infty d_n q^n \Big)   ,    
\end{align}
where $R_i$ denote  the dimensions of  irreducible representations of $\mathfrak{g}^{(\mu,\nu)}$. The coefficients $a_n, b_n, c_n$, and $d_n$   are the sum of the dimensions of the irreducible representations for the Lie algebra. The modular invariant partition function of $\mathcal{T}^{(\mu,\nu)}$ on torus is
 \begin{align}
     Z =|\chi_0|^2 + \sum_{i=1}^3 \mathcal{M}_i |\chi_i|^2, 
 \end{align}  
where $\mathcal{M}_i$ are positive integers that capture the degeneracies of primary fields. 

The first main result of this paper is the following. We find that the characters $\chi_i$ of the theory $\mathcal{T}^{(\mu,\nu)}$  satisfy the uniform fourth-order modular linear differential equation 
\begin{align}\label{MLDE1}
    [D^4+\lambda_1E_4D^2+\lambda_2E_6D+\lambda_3E_4^2]\chi=0,
\end{align}
where  the three coefficients $\lambda_1,\lambda_2$ and $\lambda_3$ are determined  by $(\mu,\nu)$ as follows:
\begin{align}
  \lambda_1&= -\frac{11 \mu^2 \nu^2+4 \mu^2 \nu+11 \mu^2+4 \mu \nu^2-28 \mu \nu+4 \mu+11 \nu^2+4 \nu+11}{36 (\mu+1)^2 (\nu+1)^2} ,\\
  \lambda_2&=\frac{49 \mu^2 \nu^2+8 \mu^2 \nu-5 \mu^2+8 \mu \nu^2-56 \mu \nu+8 \mu-5 \nu^2+8 \nu+49}{216 (\mu+1)^2 (\nu+1)^2}, \\
  \lambda_3&= -\frac{(\mu \nu-1)^2 (\mu \nu+2 \mu+1) (\mu \nu+2 \nu+1)}{16 (\mu+1)^4 (\nu+1)^4}.
\end{align} 
Specializing to $\nu=5$, the 4th order MLDE factorizes as 
\begin{align}\label{eq:factorize}
    \left(D^2-\frac{(\mu-5) (\mu+7)}{144 (\mu+1)^2}E_4\right)\left(D^2-\frac{(5 \mu-1) (7 \mu+1)}{144 (\mu+1)^2} E_4
\right)\chi=0.
\end{align}
This differential operator can thus be written as $\Delta_{1/\mu}\circ\Delta_{\mu}$ 
where $\Delta_{\mu}$ is the second order operator given in \eqref{eq:deltanu}. This factorization demonstrates that the fourth-order MLDE \eqref{MLDE1} provides a natural generalization of the second-order Mathur--Mukhi--Sen equation of the theory $\cT^{(\mu,5)}$.
Note that for $ \cT^{(\mu,5)}$ we get $h^{(\mu)}=h_3^{(\mu,5)}$ and $ \cT^{(5,\nu)}$ we get $h^{(\nu)}= h_2^{(5,\nu)}$.

It is natural to postulate that the above expressions for the central charge, conformal weights, and the modular linear differential equation \eqref{MLDE1} continue to hold for the extended magic triangle. This assumption allows us to identify two-dimensional rational conformal field theories beyond the original magic triangle of ordinary Lie groups. 
Indeed, upon specializing to $\nu=4$, we find the equation \eqref{MLDE1} reduces to a one-parameter family of fourth-order MLDEs governing some minimal $\mathcal{W}$-algebras \cite{Kawasetsu:2018tzs}. In this case, the characters of the theory $\mathcal{T}^{(\mu,4)}$ coincide with the characters of Ramond-twisted irreducible modules of the minimal $\mathcal{W}$-algebra $\mathcal{W}_{-h^\vee/6}(\mathfrak g,f_\theta)$, where $\mathfrak g$ belongs to the Cvitanovi\'c--Deligne exceptional series. 
More generally, by explicitly solving the MLDE \eqref{MLDE1} and interpreting its solutions as RCFT characters, we are led to identify a class of theories $\mathcal{T}^{(\mu,\nu)}$ extending those associated with the original magic triangle.\footnote{Note that two theories ${\mathcal T}^{(\frac23,2)}$ and $  {\mathcal T}^{(1,\frac32)}$ have different central charges $4/5$ and $6/5$ respectively, even though they have the same dimensions $0$ and dual Coxeter numbers $-1$. } 
\begin{itemize}
        \item $\cT^{(2,1)}$ is a unitary RCFT with $c=2$ and conformal weights $h_i=0,(\frac{1}{6})_6,(\frac{1}{2})_3,(\frac{2}{3})_2$. The VOA was constructed in \cite{KITAZUME200215}, see Remark 5.8 there for all its irreducible modules. This RCFT is denoted as $M_2$ in \cite[Section 5.6]{Duan:2022ltz}. The four characters have appeared as CFT-type solutions of fourth order holomorphic MLDE in \cite[Table 6]{Kaidi:2021ent}, and their   $q$-series expansions   can be found in \cite[Equation (5.34)]{Duan:2022ltz}. The $M_2$ can be realized by the coset construction $(D_4)_1/(A_2)_1$ \cite[Section 5.6]{Duan:2022ltz}. We notice that it can also be realized as a non-diagonal modular invariant of $(U_1)_2\times (U_1)_{6}$, which is consistent with magic triangle element $U_1^2$.
        
        \item $\cT^{(3/2,1)}=D_{2\rm A}$ is a unitary RCFT  associated to the 2A conjugacy class of the Monster group. It has $c=6/5$ and conformal weights $h_i=0,(\frac{1}{10})_3,(\frac{1}{2})_3,\frac{3}{5}$. The corresponding VOA can be constructed by using a code over  $\mathbb{Z}_2\times \mathbb{Z}_2$ and as an extension of the product of Ising model $\mathrm{Vir}_{4,3}$ and tri-critical Ising model $\mathrm{Vir}_{5,4}$ \cite{LAM2000268}. For 2d RCFTs associated to various conjugacy classes and sporadic groups, see a good review in \cite{Bae:2020pvv}.
        
        \item $\cT^{(3,2/3)}=\mathrm{Vir}_{6,5}^{\rm e}$ is a non-diagonal modular invariant of minimal model $\mathrm{Vir}_{6,5}$ which describes the critical 3-state Potts model. The corresponding VOA can be constructed by using a code over $\mathbb{Z}_3$ \cite{MIYAMOTO200156}. 
        \item $\cT^{(3/2,3/2)}=(A_1)_8^{\rm e}$ is the D-type modular invariant of $(A_1)_8$. The associated VOA can be constructed by simple current extension.
    \end{itemize}
All RCFTs associated with the extended magic triangle are presented in Table \ref{tb:MT}. The data of the extended magic triangle and the corresponding RCFTs are collected in the Tables \ref{tb:MTdata0}, 
\ref{tb:MTdataR}, \ref{tb:MTdata} in the Appendix \ref{sec:A}.

We now discuss the symmetry of the magic triangle at level one. It is straightforward  to verify that under the exchange of $\mu$ and $\nu$, the central charge $c$, the conformal weight $h_1$ and the fourth order MLDE are invariant, while $h_2,h_3$ are exchanged. This suggests that the $\mathcal{T}^{(\mu,\nu)}$  and $\mathcal{T}^{(\nu,\mu)}$ theories are isomorphic to each other, differing only by a relabeling of $\chi_2$ and $\chi_3$.  Consequently, it suffices to   present only half of the triangle, as shown in the Tables \ref{tb:MTdata0}, 
\ref{tb:MTdataR}, \ref{tb:MTdata}. In total, there are 30 non-equivalent RCFTs in the extended magic triangle at level one. The sixteen of these theories  arise from ordinary WZW model of semisimple Lie groups, eight arise from intermediate VOAs, three arise from Virasoro minimal models, two arise from $U_1$ theories (that is,  compact bosons on a circle), and the final theory  $D_{2\rm A}$ arises from the product of Ising model and tri-critical Ising model and  is  associated with Monster group. For all 30 theories on the magic triangle at level one, the Fourier coefficients of the characters remain non-negative integers.

Several remarks are in order:
\begin{itemize}
\item We find that the four characters of each theory $\mathcal{T}^{(\mu,\nu)}$ satisfy the following remarkable identity:
\begin{align}\label{chiid}
\chi_0\chi_1-\chi_2\chi_3=\mathrm{const}= R_1.  
\end{align} 
We refer to \eqref{chiid} as the \emph{magic bilinear relation}. Its derivation will be presented in a later section.
As an immediate consequence, this identity implies a simple relation among the dimensions $R_1,R_2,R_3$ of the corresponding representations,
\begin{align}\label{eq:RRR}
    \frac{R_2 R_3}{R_1}= \frac{12(\mu\nu-1)^2}{\mu\nu}. 
\end{align}

\item For the Cvitanovi\'c--Deligne exceptional series, there exist only two distinct characters, or even a single character in the case of $(E_8)_1$. We adopt the convention that the theory $\mathcal{T}^{(5,\nu)}$ has $\chi_0,\chi_2$ as a row, while $\cT^{(\mu,5)}$ has $\chi_0,\chi_3$ as a column. In these cases, the remaining solutions of the fourth-order MLDE exhibit Fourier coefficients with unbounded denominators and therefore cannot be interpreted as RCFT characters. Such solutions are referred to as spurious solutions. A detailed discussion of this phenomenon is deferred to Appendix~\ref{sec:B}. 

\item The thirty theories $\mathcal{T}^{(\mu,\nu)}$ identified here do not exhaust all CFT-type solutions of general fourth-order holomorphic MLDEs. For instance, the characters of the minimal model $\mathrm{Vir}_{9,2}$ also satisfy a fourth-order holomorphic MLDE. For a systematic classification of admissible solutions with no integer non-vacuum conformal weights, we refer the reader to \cite{Kaidi:2021ent}.
\end{itemize}

\subsection{Magic coset}

We find that the coset relation \eqref{E8evel1} within the Cvitanovi\'c--Deligne exceptional series is in fact a very special instance of a much broader class of coset constructions that exist among all theories in the extended magic triangle. Remarkably, for any two theories in the magic triangle one may form a coset that yields a third theory, which again belongs to the magic triangle. 
More precisely, we uncover the following universal coset relation valid throughout the extended magic triangle:
\begin{align}\label{magiccoset}
    \frac{ \mathcal{T}^{(\mu ,\rho) }}{\mathcal{T}^{(\nu ,\rho )}}=\mathcal{T}^{(\mu, 1/\nu )},\qquad \mu>\nu.
\end{align}
We refer to \eqref{magiccoset} as the \emph{magic coset}. Using the central charge formula \eqref{eq:ccharge}, it is straightforward to verify that both sides of \eqref{magiccoset} indeed have identical central charges. This relation suggests that taking the coset of any upper and lower exceptional series produces another theory within the magic triangle. As concrete examples, consider $(\mu,\nu)=(5,3)$ and $(3,2)$. The magic coset \eqref{magiccoset} yields
\begin{align}
  \frac{(\mathfrak{g}_{\rm CD})_1}{(\mathfrak{g}_{\rm sub})_1}=  (A_1)_1, \quad\textrm{and}\quad
   \frac{(\mathfrak{g}_{\rm sub})_1}{(\mathfrak{g}_{\rm Severi})_1 }=(U_1)_3 .
\end{align}
These include well-known coset constructions such as $(E_8)_1/(E_7)_1=(A_1)_1$ and $(E_7)_1/(E_6)_1=(U_1)_3$. Similarly, for $(\mu,\nu)=(5,4)$ and $(4,3)$, the magic coset gives
\begin{align}
\frac{(\mathfrak{g}_{\rm CD})_1}{(\mathfrak{g}_{\rm I})_1}= \mathrm{Vir}^{\rm eff}_{5,2},\quad\textrm{and}\quad
\frac{(\mathfrak{g}_{\rm I})_1}{(\mathfrak{g}_{\rm sub})_1}= \mathrm{Vir}^{\rm eff}_{5,3},
\end{align}
both of which were previously observed in \cite{Lee:2024fxa}. Equation \eqref{magiccoset} also indicates that by taking cosets among exceptional series one can generate every theory in the extended magic triangle at level one.

An equivalent formulation of the magic coset relation \eqref{magiccoset} is
\begin{align}\label{magiccoset2}
    \mathcal{T}^{(\mu, \nu )}=\frac{ \mathcal{T}^{(\mu ,\rho) }}{\mathcal{T}^{(1/\nu ,\rho )}}= \frac{ \mathcal{T}^{(\rho , \nu) }}{\mathcal{T}^{(\rho ,1/\mu )}} . 
\end{align} 
Setting $\rho=5$ yields
\begin{align}\label{magiccosetCD}
    \mathcal{T}^{(\mu,\nu )}=\frac{ \mathcal{T}^{(\mu ,5) }}{\mathcal{T}^{(1/\nu ,5 )}}=\frac{ \mathcal{T}^{(5,\nu ) }}{\mathcal{T}^{(5,1/\mu  )}}  . 
\end{align}
This form is particularly useful, as it implies that every theory in the extended magic triangle in Table \ref{tb:MT} can be realized as a coset between two theories belonging to the Cvitanovi\'c--Deligne exceptional series. In this sense, the magic coset construction provides a CFT analogue of the centralizer construction of Deligne--Gross for the original magic triangle \cite{DG}. 

The magic coset \eqref{magiccosetCD} first implies the following relations between the central charges
\begin{align}
    c^{(5,\nu)}& =c^{(5,1/\mu)}+ c^{(\mu,\nu)}   
    , \nonumber \\
   c^{(\mu,5)}& = c^{(\mu,\nu)} +   c^{( 1/\nu,5)}  
    ,
\end{align}
which can be easily checked with \eqref{eq:ccharge}. It further implies the following relations among the conformal weights
%
\begin{align}
    h^{(1/\mu)}+ h_3^{(\mu,\nu)}=1, &\quad \  h^{( \nu)}= h^{(\mu,\nu)}_2 =h^{(1/\mu)}+h_1^{(\mu,\nu)} , \nonumber \\ 
   h_2^{(\mu,\nu)}+h^{(1/\nu)}=1, &\quad \  h^{(\mu)}= h^{(\mu,\nu)}_3 = h_1^{(\mu,\nu)} +h^{(1/\nu)} ,    
 \end{align}
which can be checked with \eqref{eq:weight}. 
Recall the degeneracy of the non-vacuum primary field of $\mathcal{T}^{(5,\nu)}$ theory is denoted as $\mathcal{N}_{(\nu)}$, which satisfies
$   \mathcal{N}_{(\nu)}=\mathcal{N}_{(1/\nu)}.$ The numbers are collected in Table \ref{tb:CD}. 
The coset \eqref{magiccosetCD} implies that the degeneracies of $\mathcal{T}^{(\mu,\nu )}$ theory are
\begin{align}
    (\mathcal{M}^{(\mu,\nu)}_1, \mathcal{M}^{(\mu,\nu)}_2, \mathcal{M}^{(\mu,\nu)}_3)=\left(\mathcal{N}_{(\mu)}\mathcal{N}_{(\nu)},\mathcal{N}_{(\nu)},\mathcal{N}_{(\mu)}\right).
\end{align}
This shows that $\mathcal{M}_2$ remains constant along each column, while $\mathcal{M}_3$ remains constant along each row of the magic triangle at level one, as summarized in Table~\ref{tb:MTdata}. 

The magic coset \eqref{magiccosetCD} further determines the modular $S$-transformation of $\mathcal{T}^{(\mu,\nu)}$. While one might  expect the solutions of the fourth-order MLDE to form a degree-four vector-valued modular form, it is more natural in the present context to regard the four characters as forming a $2\times2$ matrix-valued modular form.\footnote{We thank Don Zagier for suggesting the interpretation in terms of matrix-valued modular forms.} Let us define the $2\times 2$ matrix $X^{(\mu,\nu)}$ for the characters  of $\cT^{(\mu,\nu)}$ theory as
\begin{equation}
  X^{(\mu,\nu)}=\begin{pmatrix}
\chi_0 & \chi_2 \\
\chi_3 & \chi_1 
\end{pmatrix}.  
\end{equation}
We then find the following elegant $S$-transformation law:
\begin{align}\label{eq:S22}
X^{(\mu,\nu)}(-1/\tau) = S_{(\mu)}\cdot  X^{(\mu,\nu)}(\tau)\cdot S_{(\nu)}^T, 
    \end{align}
where $S_{(\nu)}$ denotes the $2\times2$ $S$-matrix of the theory $\mathcal{T}^{(5,\nu)}$ in the Cvitanovi\'c--Deligne exceptional series,  given explicitly in \ref{sec:B0}.  
Taking the determinant of \eqref{eq:S22} and using the fact $\det S_{(\nu)}=-1$, we see that $\det(X^{(\mu,\nu)})=\chi_0 \chi_1 - \chi_2\chi_3$ is modular invariant. Moreover, since $ \alpha_0+\alpha_1=0 $ and $\alpha_2+\alpha_3=1$, this determinant is holomorphic and must therefore be constant. A direct computation shows that the constant equals $R_1$, thereby proving the magic bilinear relation \eqref{chiid}.

The magic coset construction also leads to a variety of nontrivial relations among characters of level-one theories in the extended magic triangle. In particular, \eqref{magiccosetCD} implies the following relations between the characters of the Cvitanovi\'c--Deligne exceptional series and those of a general theory $\mathcal{T}^{(\mu,\nu)}$:
\begin{align}
    \left(\begin{array}{lr}\chi_0 &    \chi_2 \end{array}\right)^{(5, \nu)} & =  \left(\begin{array}{cc}\chi_0 &  \chi_2 \end{array}\right)^{(5,1/\mu)} \cdot \left(\begin{array}{cc}1 &  0 \\ 0 & \cN_{(\mu)} \end{array}\right)  \cdot   X^{(\mu,\nu)}, \label{eq:Qu01} \\
    \left(\begin{array}{c}\chi_0 \\  \chi_3 \end{array}\right)^{(\mu,5)} & =    X^{(\mu,\nu)} \cdot \left(\begin{array}{cc}1 &  0 \\ 0 & \cN_{(\nu)} \end{array}\right)  \cdot 
    \left(\begin{array}{c}\chi_0 \\  \chi_3 \end{array}\right)^{(1/\nu,5)} .  \label{eq:Qu02}
\end{align}
Extracting the leading terms in $q$-expansion yields the following relations among representation dimensions:
%
\begin{align}
 & R_1^{(\mu,\nu)} = \frac{\mu\nu}{12(5\mu-1)(5\nu-1)} R_{(\mu)}R_{(\nu)}, \label{dimR1}\\
 & R_2^{(\mu,\nu)} = \frac{\mu\nu-1}{  5\nu-1}  R_{(\nu)}, \label{dimR2} \\
 & R_3^{(\mu,\nu)} = \frac{\mu\nu-1}{ 5\mu-1 } R_{(\mu)}. \label{dimR3}
\end{align} 
Here $R_{(\nu)}$ denotes the dimension of the fundamental representation of the Cvitanović--Deligne exceptional series, cf. \eqref{eq:CDchi} and Table \ref{tb:CD}. These relations are easily checked to be consistent with \eqref{eq:RRR}.
For the row $\mathcal{T}^{(5,\nu)}$, the expression for $R_3^{(\mu,\nu)}$ becomes linear in $\nu$ and takes the form
\begin{align} & \Eh:  12\nu-3, \quad E_7: 12\nu-4, \quad E_6:  6\nu-3, \quad F_4:  6\nu-4, \\
 & D_4:  2\nu-2, \quad G_2: 2\nu-3, \quad A_2:  \nu -2, \quad A_1:  \nu-3, \quad A_{1/2}:  \nu-4. \end{align}
This precisely reproduces the extension of the preferred representation dimensions found by Deligne--Gross \cite[Equation 2]{DG}. 

More generally, the magic coset relation \eqref{magiccoset2} implies the following character identity among three theories in the magic triangle:
\begin{align}\label{eq:3X}
    X^{(\mu,\nu)} & =  X^{(\mu,\rho)}  \left(\begin{array}{cc} 1 & 0 \\ 0 & \cN_{(\rho)} \end{array} \right)  X^{(1/\rho,\nu)} ,  
\end{align} 
for $\rho\le \nu$ and $1/\rho\le  \mu$. 
At leading order, this yields the following relations among representation dimensions:
\begin{align}
        d_\mathfrak{g}^{(\mu,\nu)}&=   d_\mathfrak{g}^{(\mu,\rho)} + d_\mathfrak{g}^{(1/\rho,\nu)}+\cN_{(\rho)} R_2^{(\mu, \rho)} R_3^{(1/\rho,\nu)},  \\
  R_1^{(\mu,\nu)}&= \cN_{(\rho)} R_1^{(\mu,\rho)}R_1^{(1/\rho, \nu)} ,    \\ 
  R_2^{(\mu,\nu)}&=  R_2^{(1/\rho,\nu)}  +\cN_{(\rho)}R_2^{(\mu,\rho)}R_1^{(1/\rho, \nu)} ,     \\ 
R_3^{(\mu,\nu)}&=   R_3^{(\mu, \rho)} +\cN_{(\rho)} R_1^{(\mu,\rho)}R_3^{(1/\rho, \nu)}.  
\end{align}

\subsection{Five atomic theories}

An interesting consequence of the magic coset relation \eqref{magiccoset} is that all thirty CFTs appearing in the extended magic triangle in Table~\ref{tb:MT} can be generated from only \emph{five atomic theories} lying just below the line $\mu\nu=1$. These five atoms are 
\begin{align}
    \mathrm{Vir}_{5,2}^{\rm eff},\ \mathrm{Vir}_{5,3}^{\rm eff},\ (U_1)_3,\ \mathrm{Vir}_{6,5}^{\rm e},\ D_{2\rm A}.
\end{align}
This means that every theory $\mathcal{T}^{(\mu,\nu)}$ can be realized as an extension of a product of these atomic theories, or equivalently as a non-diagonal modular invariant of an  tensor product CFT of the atoms. More concretely, given a theory $\mathcal{T}^{(\mu,\nu)}$ in Table~\ref{tb:MT}, we draw a right triangle bounded by the three lines $\mu=\mathrm{const}$, $\nu=\mathrm{const}$, and $\mu\nu=1$. The theory $\mathcal{T}^{(\mu,\nu)}$ with $\mu\nu>1$ is then constructed from the atomic theories that lie along the segment of the slope just below the line $\mu\nu=1$ within this triangle. Since each of the five atoms appears twice just below   the $\mu\nu=1$ line, any theory in the extended magic triangle involves at most two identical atomic constituents. The central charge of the $\cT^{(\mu,\nu)}$ is just the sum of the effective central charges of all constituent atoms, which are positive. The conformal weights and characters of atoms are weaved more delicately to form the theory $T^{(\mu,\nu)}$. The basic data of these five atoms are summarized in Table~\ref{tab:atom}.

\begin{table}[h]
     \centering
     \begin{tabular}{|c|c|c|c|c|c|} \hline
        $\mathcal{T}^{(\mu,\nu)}$ & Vir$^{\rm eff}_{5,2}$ & Vir$^{\rm eff}_{5,3}$ & $(U_1)_3$ & Vir$^{\rm e}_{6,5}$ & $D_{\rm 2A}$\\ \hline
       $(\mu,\nu)$  & $(1/4,5)$ & $(1/3,4)$ & $(1/2,3)$ & $(2/3,2)$ & $(1,3/2)$ \\ \hline
       $\dim$ & 1 & 1 & 1 & 0 & 0 \\ \hline
       $h^\vee $ & $3/2$ &$ 2/3$ & 0 &$ -1$ & $-1 $\\ \hline
       $c$ & 2/5 & 3/5 & 1 & 4/5 & 6/5 \\ \hline
      $h_1, R_1 $ & $-  $ & $1/20,1 $ & $(1/12)_2,1 $ & $(1/15)_2,1 $ & $(1/10)_3,1 $ \\ \hline
      $h_2, R_2$ & $1/5,1$ & $1/4,1$ & $(1/3)_2,1 $ & $2/5,1 $ & $(1/2)_3,1 $ \\ \hline
      $h_3, R_3$ & $- $ & $4/5,1$ & $3/4,2 $ & $(2/3)_2,1 $ & $3/5,2$ \\ \hline
     \end{tabular}
     \caption{The five atoms.}
     \label{tab:atom}
 \end{table}

We first illustrate this construction for the Cvitanovi\'c--Deligne exceptional series at level one using the coset relation \eqref{magiccosetCD}. Starting from the bottom-left corner of Table~\ref{tb:MT}, we observe that
$\cT^{(5,1/4)}={\rm Vir}_{5,2}^{\rm eff}$ is itself one of the atomic theories. The next theory, $\cT^{(5,1/3)}=(A_1)_1$ WZW model is obtained by adjoining $\cT^{(4,1/3)}={\rm Vir}_{5,3}^{\rm eff}$ to ${\rm Vir}_{5,2}^{\rm eff}$. Similarly, the third theory, $\cT^{(5,1/2)}=(A_2)_1$ WZW model is obtained by adjoining $\cT^{(3,1/2)}=(U_1)_3$ to $(A_1)_1$. 
Proceeding in this manner, one successively constructs the level-one WZW models of the Cvitanovi\'c--Deligne exceptional series by accumulating diagonal elements. In particular, combining the first five atoms,   from $\cT^{(5,1/4)}={\rm Vir}_{5,2}^{\rm eff}$ to $\cT^{(3/2,1)}=D_{2\rm A}$  yields the $(D_4)_1$ WZW model, while taking two copies of each of these five atoms produces the $(E_8)_1$ WZW model. As expected, the sum of the central charges of these atoms is eight as $2(2/5+3/5+1+4/5+6/5)=8.$

In summary, any theory $\mathcal{T}^{(\mu,\nu)}$ listed in Table~\ref{tb:MT} can be obtained by combining all atomic theories lying along the skew diagonal that runs from the leftmost atom in the $\mu$-th row to the uppermost atom in the $\nu$-th column. For instance,  $\cT^{(1,2)}=M_2$ is constructed from ${\rm Vir}^{\rm e}_{6,5}$ and $D_{2\rm A}$. The $\cT^{(3,3/2)} = (C_3)_1$ is built of $(U_1)_3, {\rm Vir}_{6,5}^{\rm e}$ and two  $D_{2\rm A}$'s.

\subsection{The diagonal exceptional series}
A distinguished specialization of the magic coset relation \eqref{magiccoset} is obtained by setting $\rho=\mu$, which yields
\begin{align}\label{magiccoset3}
   \mathcal{T}^{(\mu ,1/\nu) }=\frac{ \mathcal{T}^{(\mu ,\mu) }}{\mathcal{T}^{(\mu ,\nu) }},\qquad \mu>\nu.
\end{align}
This relation provides a direct generalization of the $(E_8)_1$ dual pairs appearing in \eqref{E8evel1}. For instance, when $\mu=3$, equation \eqref{magiccoset3} implies that the RCFTs in the subexceptional series at level one naturally pair via coset constructions with respect to $(D_6)_1$. Given the special interest on $\mathcal{T}^{(\mu ,\mu) }$ theories, namely,
$$
(A_1)_8^{\rm e} ,\  (A_2)_1^2 ,\ (D_6)_1,\ (\Dhh)_1,\ (E_8)_1,
$$
we refer to them as the \emph{diagonal exceptional series}. Each of these theories admits a $\mathbb{Z}_2$ inner automorphism such that $h_2=h_3$. Consequently, the theory $\mathcal{T}^{(\mu,\mu)}$ possesses three independent characters with conformal weights $h_i=0,$  $( \mu-1)/ (\mu+1)$, $ \mu/(\mu+1)$. These characters satisfy a third-order holomorphic modular linear differential equation of the form
\begin{equation}
    [D^3 +\lambda_1 E_4 D+\lambda_2 E_6]\chi=0,
\end{equation}  
with
\begin{equation}
    \lambda_1= -\frac{11 \mu^2-14 \mu+11}{36 (\mu+1)^2}, \quad \lambda_2 =\frac{(\mu-1)^2}{8(\mu+1)^2}. 
\end{equation}

The relation \eqref{magiccoset3} gives rise to a variety of nontrivial coset constructions. Here we record two particularly simple examples, which to our knowledge have not appeared previously in the literature:
\begin{align}\label{d2acoset}
    D_{2\rm A} & =\frac{(A_1)_8^{\rm e} }{D_{2\rm A} },\\
    M_2 &=\frac{(A_2)_1^2}{M_2}.\label{m2coset}
\end{align}
Using the general character identity \eqref{eq:3X}, one readily derives the corresponding character relations. For the coset \eqref{d2acoset}, one finds
\begin{align}
  \chi^{(A_1)_8^{\rm e}}_0= \chi^{(A_1)_8}_0 +\chi^{(A_1)_8}_{2}=\left(\chi^{D_{2\rm A}}_0\right)^2 +3\Big(\chi^{D_{2\rm A}}_{\frac12}\Big)^2  .
\end{align}
Here, and in what follows, the subscript of $\chi$ denotes the conformal weight of the corresponding character.
Similarly, for the coset \eqref{m2coset}, the character relation takes the form
\begin{align}
    \left(\chi^{(A_2)_1}_0\right)^2 & =\left(\chi^{M_2}_0\right)^2+3\Big(\chi^{M_2}_{\frac12}\Big)^2 .
\end{align}

\section{Magic triangle at level two} \label{sec:4}
At level two, the magic triangle no longer behaves as perfectly as at level one. For the 18 elements that belong to the original magic triangle in Table~\ref{tb:MT0}, the associated higher-level theories do not present conceptual difficulties, and the corresponding RCFTs are well defined. By contrast, the remaining 12 theories in the extended triangle require more careful treatment at higher levels.
On the one hand, the two elements with vanishing dimension $d=0$ and dual Coxeter number $h^\vee=-1$ become trivial at higher levels, since their central charges vanish identically. On the other hand, some intermediate Lie algebras do not admit consistent higher-level generalizations. A typical example is the case $E_{7\frac12}$, for which we have argued that only levels $k\leq 5$ can be realized as RCFTs \cite{Lee:2023owa}. 

Using the dimension and dual Coxeter number formulas of the magic  triangle, one obtains a universal expression for the level-two central charge,
\begin{align}\label{eq:c2}
c_2= \frac{24(\mu\nu-1)(\mu\nu+\mu+\nu-4)}{(\mu+1)(\nu+1)(\mu\nu+\mu+\nu-3)}.
\end{align}
In principle, all 13 conformal weights can be written explicitly as functions of $\mu$ and $\nu$. In particular, there is always a  character with exponent zero. 
We expect that the characters of all level-two theories satisfy a two-parameter family of holomorphic modular linear differential equations of order 13 with index $l=36$. Owing to their complexity, we do not present these equations explicitly. Nevertheless, substantial simplifications occur for certain exceptional series. For example, the characters of the CD exceptional series at level two satisfy a one-parameter family of sixth-order holomorphic MLDEs \cite{Lee:2023owa}. Moreover, the intermediate exceptional series at level two are realized by minimal $\mathcal{W}$-algebras  $\mathcal{W}_{-h^\vee/6+1}(\mg_{\rm CD},f_\theta)$, which has been studied in \cite{AM,Sun:2024mfz,arakawa2024lisse,Kac:2024}. 
Finally, for the subexceptional series at level two, an emergent supersymmetry plays a crucial role. In this case, it is most natural to fermionize the theories and study their fermionic characters and fermionic MLDEs. A detailed analysis of this structure will be presented in the next subsection. Finally, we expect that the Severi exceptional series at level two should admit a $\mathbb{Z}_3$ parafermionic structure. It would be interesting to investigate this in the future. 

\subsection{Subexceptional series at level two}
A remarkable phenomenon we observe is that all theories in subexceptional series at level two are supersymmetric. These theories $\cT_2^{(3,\nu)}$ are
\begin{equation}\label{eq:sub2}
    (U_1)_6,\ (A_1)_6,\ (A_1^{\oplus 3})_2,\ (C_3)_2,\ (A_5)_2,\ (D_6)_2,\ (\Dh)_2,\ (E_7)_2.
\end{equation}
A prerequisite of emergent supersymmetry is that the RCFT has a primary field with conformal weight $3/2$, which serves as the supersymmetry generator. It is easy to verify this is true for all theories in \eqref{eq:c2}. In many cases, the supersymmetry of these theories has been studied extensively, see e.g. \cite{Johnson-Freyd:2019wgb,Bae:2021lvk}.  In particular, the $(\Dh)_2$ was found to be supersymmetric only recently in \cite{Lee:2024fxa}. setting $\mu=3$ in \eqref{eq:c2}, the central charge of $\cT_2^{(3,\nu)}$ theory  becomes
\begin{align}
 c=   \frac{3(3 \nu-1) (4 \nu-1)}{2 \nu (\nu+1)}.
\end{align}

We observe that the fermionization of each theory in \eqref{eq:sub2}, denoted as $\cF \cT_2^{(3,\nu)}$, has at most four distinct Neveu--Schwarz characters and an equal number of Ramond characters. In general, we find that  the four Neveu--Schwarz conformal weights are
\begin{align}\label{eq:NSweight}
  0,\frac{4 \nu-1}{4 (\nu+1)},\frac{2 \nu-1}{2 \nu},\frac{7 \nu^2-7 \nu+6}{4 \nu (\nu+1)}
 ,
\end{align}
while the four Ramond conformal weights are
\begin{align}\label{eq:Rweight}
\frac{(3 \nu-1) (4 \nu-1)}{16 \nu (\nu+1)},\frac{3 (4 \nu-1)}{16 \nu},\frac{(3 \nu-1) (8 \nu+3)}{16 \nu (\nu+1)},\frac{3 (8 \nu-5)}{16 \nu}
.
\end{align}
Notice the first Ramond weight has exponent $0$ which indicates the existence of the constant $\widetilde{\rm R}$ character. One can also easily recover 13 bosonic conformal weights and verify  that the 13 conformal weights always yield a 13th order MLDE with index $l=36$. For theories with less than four distinct NS characters, the fermionic weights take values in the  list above. 

We further find a one-parameter family of holomorphic fermionic MLDE of degree four for the Neveu--Schwarz characters of the theories $\cF \cT_2^{(3,\nu)}$.  
To present the fermionic MLDE, 
we first recall the Jacobi theta functions
\begin{align}
    \theta_2=\sum_{n\in\ZZ}q^{(n+\frac12)^2/2},\quad \theta_3=\sum_{n\in\ZZ}q^{n^2/2},\quad \theta_4=\sum_{n\in\ZZ}(-1)^nq^{n^2/2}.
\end{align}
The most general form of fermionic MLDE with degree four and index $l=0$ for Neveu--Schwarz characters is \cite{Duan:2022kxr}
\begin{align}
      \Big[
  &  D^4  + \lambda_1(\theta_4^4 - \theta_2^4) D^3 + \big(\lambda_2 (\theta_2^8 + \theta_4^8) + \lambda_3 \theta_2^4 \theta_4^4\big)  {D}^2 
    + \big( \lambda_4(\theta_4^{8}\theta_2^4 - \theta_2^{8}\theta_4^4)
    \\ \nonumber
     & + \lambda_5 (\theta_4^{12} - \theta_2^{12}) \big) D + \lambda_6 (\theta_2^{16} + \theta_4^{16}) + \lambda_7(\theta_4^{12}\theta_2^4 + \theta_2^{12}\theta_4^4) + \lambda_8 \theta_2^8 \theta_4^8 \Big]\chi^{\rm NS}=0.
\end{align}
In the meantime, the Ramond characters satisfy the following dual MLDE obtained from modular transformation
\begin{align}
    \Big[ 
    D^4 & - \lambda_1(\theta_3^4 + \theta_4^4)D^3 + \big(\lambda_2 (\theta_4^8 + \theta_3^8) - \lambda_3 \theta_4^4 \theta_3^4\big )D^2 + \big( \lambda_4 (\theta_3^{8}\theta_4^4 + \theta_4^{8}\theta_3^4)  
    \\ \nonumber
     -&\, \lambda_5 (\theta_3^{12} + \theta_4^{12})\big) D 
    + \lambda_6 (\theta_4^{16} + \theta_3^{16}) - \lambda_7(\theta_3^{12}\theta_4^4 + \theta_4^{12}\theta_3^4) + \lambda_8 \theta_4^8 \theta_3^8 \Big]\chi^{\rm R} =0.   
\end{align}
The eight fermionic weights \eqref{eq:NSweight} and \eqref{eq:Rweight} along with the no free fermion condition allow us to solve the unfixed coefficients $\lambda_i$ as rational functions of $\nu$. In the end, we find the one-parameter family given by
\begin{align}\label{eq:mu8}
   & \lambda_1= \frac{\nu^2+3 \nu-3}{4 \nu (\nu+1)},\ \ \lambda_{2}= -\frac{856 \nu^4-1654 \nu^3+2827 \nu^2-2610 \nu+1053}{1152 \nu^2 (\nu+1)^2},\\ \nonumber
   &\lambda_{3}= -\frac{860 \nu^4-1070 \nu^3+3047 \nu^2-3546 \nu+1377}{576 \nu^2 (\nu+1)^2},\\ \nonumber
   &\lambda_{4}=\frac{2560 \nu^6+1488 \nu^5-13605 \nu^4+22117 \nu^3-14274 \nu^2+4401 \nu-675}{9216 \nu^3 (\nu+1)^3},\\\nonumber
   &\lambda_{5}= \frac{6896 \nu^6+6816 \nu^5-33111 \nu^4+60047 \nu^3-48078 \nu^2+17955 \nu-2889}{27648 \nu^3 (\nu+1)^3},\\\nonumber
   &\lambda_{6}= -\frac{(3 \nu-1) (4 \nu-1)^2 \left(4 \nu^2+15 \nu-9\right) \left(16 \nu^2-21 \nu+23\right)}{65536 \nu^4 (\nu+1)^3},\\\nonumber
   &\lambda_{7}= \frac{3 (3 \nu-1) (4 \nu-1) \left(32 \nu^6-740 \nu^5+1740 \nu^4-2133 \nu^3+1121 \nu^2-207 \nu+27\right)}{16384 \nu^4 (\nu+1)^4}, \\ \nonumber 
 & \lambda_8=  2\lambda_7-2\lambda_6.
\end{align}
 The last equation comes from the supersymmetry condition that one Ramond conformal weight has zero exponent.  We have explicitly checked  that the Neveu--Schwarz and Ramond characters of the theories  $\cF \cT_2^{(3,\nu)}$   indeed satisfy the above fermionic MLDEs. This one-parameter family of fermionic MLDE suggests that all Neveu--Schwarz and Ramond characters of the theories $\cF \cT_2^{(3,\nu)}$   can be    solved uniformly  as functions of $\nu$.

\subsection{Coset constructions}
In contrast to the level-one case, there no longer exists a universal coset construction covering the entire magic triangle at level two. Nevertheless, the magic coset relation \eqref{magiccoset} continues to provide strong guidance and motivates us to search for new uniform coset constructions involving the Cvitanovi\'c--Deligne, subexceptional, and Severi exceptional series at level two. 

We first find the following universal coset construction between the Cvitanovi\'c--Deligne and subexceptional series at level two:
\begin{align}\label{coset2}
\frac{\mathcal{T}^{(5,\nu)}_2}{\mathcal{T}^{(3,\nu)}_2\times \mathcal{T}^{(5,1/3)}_2}=    \frac{(\mathfrak{g}_{\rm CD})_2}{(\mathfrak{g}_{\rm sub})_2\times (A_1)_2}=\mathrm{Vir}_{n+1,n},\qquad n=
    \left\{  
             \begin{array}{ll}  
             3+\frac{4}{\nu-1},\ & \nu>1 \\  
             \frac{4}{1-\nu}-4,\ \ & \nu<1.   
             \end{array}  
\right.  
\end{align}
This coset holds directly for $\nu= \frac12,\frac23,\frac32,2,3,5$. For $\nu=\frac14,4$, the relation continues to hold after rescaling $n$ and $n+1$ so that they are coprime integers, in which case both sides of the coset become non-unitary. 
To the best of our knowledge, the uniform coset construction \eqref{coset2}, as well as the one in \eqref{coset2sub} below, are new. 
As an illustrative example, for $\nu=\frac23$, the coset \eqref{coset2} yields
\begin{align}
    \frac{(G_2)_2}{(A_1)_6\times (A_1)_2}=\mathrm{Vir}_{9,8}.
\end{align}
Each of the four characters of $(G_2)_2$ decomposes into the sum of 12 terms of  $(A_1)_6\times (A_1)_2\times \mathrm{Vir}_{9,8}$. For instance, denote $\mathrm{Vir}_{9,8}$ as $M$.  For the vacuum character, we have
\begin{align}
    \chi^{(G_2)_2}_0=&\ \chi^{(A_1)_2}_{0}\chi^{(A_1)_6}_{0}\chi^{M}_0+\chi^{(A_1)_2}_{\frac{3}{16}}\chi^{(A_1)_6}_{\frac{15}{32}}\chi^{M}_{\frac{11}{32}} + \chi^{(A_1)_2}_{0}\chi^{(A_1)_6}_{\frac{3}{4}}\chi^{M}_{\frac{5}{4}} \\\nonumber &+ \chi^{(A_1)_2}_{\frac12}\chi^{(A_1)_6}_{\frac{1}{4}}\chi^{M}_{\frac{5}{4}} + \chi^{(A_1)_2}_{\frac12}\chi^{(A_1)_6}_{\frac{3}{2}}\chi^{M}_{0} + \chi^{(A_1)_2}_{0}\chi^{(A_1)_6}_{\frac{3}{4}}\chi^{M}_{\frac{5}{4}} \\\nonumber
    &+ \chi^{(A_1)_2}_{\frac{3}{16}}\chi^{(A_1)_6}_{\frac{3}{32}}\chi^{M}_{\frac{87}{32}} + \chi^{(A_1)_2}_{\frac{3}{16}}\chi^{(A_1)_6}_{\frac{35}{32}}\chi^{M}_{\frac{87}{32}} + \chi^{(A_1)_2}_{0}\chi^{(A_1)_6}_{\frac{1}{4}}\chi^{M}_{\frac{19}{4}}\\ \nonumber
    &+ \chi^{(A_1)_2}_{\frac{1}{2}}\chi^{(A_1)_6}_{\frac{3}{4}}\chi^{M}_{\frac{19}{4}} + \chi^{(A_1)_2}_{\frac{3}{16}}\chi^{(A_1)_6}_{\frac{15}{32}}\chi^{M}_{\frac{235}{32}} + \chi^{(A_1)_2}_{0}\chi^{(A_1)_6}_{\frac{3}{2}}\chi^{M}_{\frac{21}{2}} .
\end{align}
The twelve terms are ordered according to increasing total conformal weight.

We further find the following universal coset construction between the subexceptional and Severi exceptional series at level 2:
\begin{align}\label{coset2sub}
\frac{\mathcal{T}^{(3,\nu)}_2}{\mathcal{T}^{(2,\nu)}_2\times \mathcal{T}^{(3,1/2)}_2}=     \frac{(\mathfrak{g}_{\rm sub})_2}{(\mathfrak{g}_{\rm Severi})_2\times (U_1)_6}=\mathcal{B}\mathrm{Vir}_{n+2,n}^{N=1},\qquad n=
    \left\{  
             \begin{array}{ll}  
             4+\frac{4}{\nu-1},\ & \nu>1, \\  
             \frac{4}{1-\nu}-6,\ \ & \nu<1.   
             \end{array}  
\right.  
\end{align}
Here again $\nu=\frac12,\frac23,\frac32,2,3,5$. The theory  $\mathcal{B}\mathrm{Vir}_{n+2,n}^{N=1}$ denotes the bosonization $\mathcal{B}$ of the unitary $N=1$ supersymmetric minimal model  $\mathrm{Vir}_{n+2,n}^{N=1}$. 
As an example for   $\nu=3$ the coset (\ref{coset2sub}) gives
\begin{align}
    \frac{(E_7)_2}{(E_6)_2\times (U_1)_6}=\mathcal{B}\mathrm{Vir}_{7,5}^{N=1}
\end{align}
It is straightforward to verify that  find $\mathcal{B}\mathrm{Vir}_{7,5}^{N=1}$ has in total $18$ bosonic primary fields with conformal weights
$$
\left\{0,\frac{3}{70},\frac{29}{560},\frac{9}{112},\frac{4}{35},\frac{3}{14},\frac{27}{80},\frac{269}{560},\frac{19}{35},\frac{43}{70},\frac{73}{112},\frac{5}{7},\frac{9}{10},\frac{8}{7},\frac{7}{5},\frac{3}{2},\frac{23}{14},\frac{31}{16}\right\}.
$$
Denote $\mathcal{B}\mathrm{Vir}_{7,5}^{N=1}$ as $\mathcal{B}$,   the vacuum  character satisfies the following relation:
\begin{align}
    \chi^{(E_7)_2}_0=&\ \chi^{(E_6)_2}_{0}\chi^{(U_1)_6}_{0}\chi^{\mathcal{B}}_0+2\chi^{(E_6)_2}_{\frac{13}{21} }\chi^{(U_1)_6}_{ \frac16}\chi^{\mathcal{B}}_{\frac{3}{14} }+\chi^{(E_6)_2}_{\frac67 }\chi^{(U_1)_6}_{ 0}\chi^{\mathcal{B}}_{\frac87 }\\ \nonumber
    &+\chi^{(E_6)_2}_{ \frac97}\chi^{(U_1)_6}_{0 }\chi^{\mathcal{B}}_{ \frac57}+2\chi^{(E_6)_2}_{ \frac{13}{21}}\chi^{(U_1)_6}_{\frac23 }\chi^{\mathcal{B}}_{ \frac57}+2\chi^{(E_6)_2}_{\frac{4}{3} }\chi^{(U_1)_6}_{\frac23 }\chi^{\mathcal{B}}_{0 }\\ \nonumber&+2\chi^{(E_6)_2}_{\frac{25}{21} }\chi^{(U_1)_6}_{ \frac16}\chi^{\mathcal{B}}_{ \frac{23}{14}}+2\chi^{(E_6)_2}_{ \frac43}\chi^{(U_1)_6}_{ \frac16}\chi^{\mathcal{B}}_{\frac32 }+2\chi^{(E_6)_2}_{\frac{25}{21} }\chi^{(U_1)_6}_{ \frac23}\chi^{\mathcal{B}}_{\frac87 }\\ \nonumber&+\chi^{(E_6)_2}_{ 0}\chi^{(U_1)_6}_{ \frac32}\chi^{\mathcal{B}}_{ \frac32}+\chi^{(E_6)_2}_{ \frac97}\chi^{(U_1)_6}_{ \frac32}\chi^{\mathcal{B}}_{ \frac{3}{14}}+\chi^{(E_6)_2}_{\frac67 }\chi^{(U_1)_6}_{\frac32 }\chi^{\mathcal{B}}_{\frac{23}{14} } .
\end{align}
We have checked that the degeneracies match,  and it is straightforward to upgrade the character relation to module decomposition.

Furthermore, there exists the following universal coset \cite{Cheng:2020srs} for Cvitanovi\'c--Deligne exceptional series resembling the Maverick coset \cite{Dunbar:1993hr}
\begin{align}\label{coset4}
 \frac{\mathcal{T}^{(5,\nu)}_1\times \mathcal{T}^{(5,\nu)}_1}{\mathcal{T}^{(5,\nu)}_2}=   \frac{(\mathfrak{g}_{\rm CD})_1\times (\mathfrak{g}_{\rm CD})_1}{(\mathfrak{g}_{\rm CD})_2} =\mathrm{Vir}_{n+1,n},\qquad n=
    \left\{  
             \begin{array}{ll}  
             2+\frac{4}{\nu-1}, & \nu>1 \\  
             \frac{4}{1-\nu}-3,\ \ & \nu<1.   
             \end{array}  
\right.  
\end{align}
Here   $\nu= \frac12,\frac23,\frac32,2,3,5$. For $\nu=\frac14,4$, the relation continues to hold  provided   one scales $n$ and $n+1$ so that they are coprime integers and choose the appropriate non-diagonal modular invariant.  For $\nu=4$, such a non-diagonal modular invariant was identified in \cite{Lee:2023owa}. For $\nu=\frac14$, the corresponding coset was recently proved in \cite{feng2025structure},  with $\mathfrak{g}^{(5,1/2)}$ as $osp(1|2)$. 
We further  find a similar coset construction for the subexceptional series:
\begin{align}\label{coset5}
\frac{\mathcal{T}^{(3,\nu)}_1\times \mathcal{T}^{(3,\nu)}_1}{\mathcal{T}^{(3,\nu)}_2}=     \frac{(\mathfrak{g}_{\rm sub})_1\times (\mathfrak{g}_{\rm sub})_1}{(\mathfrak{g}_{\rm sub})_2} =\mathcal{B}\mathrm{Vir}_{n+2,n}^{N=1},\qquad n=
    \left\{  
             \begin{array}{ll}  
             2+\frac{4}{\nu-1},\ & \nu>1, \\  
             \frac{4}{1-\nu}-4,\ \ & \nu<1. 
             \end{array}  
\right.  
\end{align}

Finally, we  make a remark concerning the intermediate exceptional series at level 2. For $\nu=2,3,5$ and $n=3+\frac{4}{\nu-1}$, the following cosets were found in \cite[Equation (1.9),(1.10)]{Sun:2024mfz}:
\begin{align}\label{coset6}
\frac{\mathcal{T}^{(5,\nu)}_2}{\mathcal{T}^{(4,\nu)}_2}&=    \frac{(\mathfrak{g}_{\rm CD})_2}{(\mathfrak{g}_{I})_2}=(\mathrm{Vir}_{ 3n+1,n })^{
\rm eff
},\\
\label{coset7}
\frac{\mathcal{T}^{(4,\nu)}_2}{\mathcal{T}^{(3,\nu)}_2}&=    \frac{(\mathfrak{g}_{I})_2}{(\mathfrak{g}_{\rm sub})_2}=(\mathcal{B}\mathrm{Vir}^{N=1}_{3n+1 , n+1})^{\rm eff}. 
\end{align}

\section{Magic triangle at arbitrary positive integer levels}\label{sec:5}
We denote the RCFT associated with the triangle element $(\mu,\nu)$ at level $k$ as $\mathcal{T}^{(\mu,\nu)}_k$. For the 16 elements of the magic triangle that are semisimple Lie algebras $\mg$, the corresponding RCFT with arbitrary positive integer level $k$ are just their WZW models $(\mg)_k$, except for ${\mathfrak{g}}^{(3,2/3)}=A_1$, ${\mathfrak{g}}^{(3/2,3/2)}=A_1$ and ${\mathfrak{g}}^{(2,3/2)}=A_2$ for which the levels must be rescaled by factors   $3$, $8$, and $2$, respectively. For the remaining 14 elements, a case-by-case analysis is required. By carefully examining the central charges and conformal weights, we arrive at the following conclusions: 
\begin{itemize}
\item $\mathcal{T}^{(1/4,5)}_k$ is the effective minimal model $\mathrm{Vir}_{2k+3,2}^{\rm eff}$, see e.g. \cite{Duan:2022ltz}. This is denoted as level-$k$ Lee-Yang model $(LY)_k$ in \cite{Duan:2022ltz}. 
\item $\mathcal{T}^{(1/3,4)}_k$ is the effective minimal model $\mathrm{Vir}_{3k+2,3}^{\rm eff}$. This was realized in \cite{Lee:2024fxa}.
\item $\mathcal{T}^{(1/2,3)}_k$ is the compact boson $(U_1)_{3k}$ with $c=1$. The central charge is due to $d_{\mathfrak{g}}=1$ and  $h^\vee=0$.
\item $\mathcal{T}^{(2/3,2)}_k$ and $\mathcal{T}^{(1,3/2)}_k$ are trivial for $k>1$ due to $d =0$ and $h^\vee =-1$. 
\item $\mathcal{T}^{(1,2)}_k$ is a non-diagonal modular invariant of $(U_1)_{2k}\times (U_1)_{6k}$ with $c=2$. The central charge is due to $d_\mathfrak{g}=2$ and $h^\vee=0$.    The vacuum character is given by
    \begin{align}
      \chi^{(U_1)_{2k}}_0\chi^{(U_1)_{6k} }_0+ \chi^{(U_1)_{2k}}_{\frac{k}{2}}\chi^{(U_1)_{6k} }_{\frac{3k}{2}} .
    \end{align}
\item  $\mathcal{T}^{(3/2,3/2)}_k$ is the $D$-type non-diagonal modular invariant of $(A_1)_{8k}$.

\item The theories $\mathcal{T}^{(4,5)}_k$, $\mathcal{T}^{(4,3)}_k$, and  $\mathcal{T}^{(4,2)}_k$ exist up to $k=5,3,2$ respectively. These theories are equivalent to the minimal $\mathcal{W}$-algebras of E-type with negative integer levels, which have recently been studied in \cite{AM,Sun:2024mfz,arakawa2024lisse,Kac:2024}. Other $\mathcal{T}^{(4,\nu)}_k$ theories for $k>1$ are not expected to exist.
\end{itemize}

\section{Concluding remarks}

In this work, we have extended the magic triangle and identified the two-dimensional conformal field theories associated with its elements at arbitrary positive integer levels. In particular, for level-one theories we uncover a number of remarkable  properties, including modular linear differential equations, the magic coset, five atomic theories, magic bilinear relations,  the dimensions and degeneracies of primary operators,     and matrix-valued modularity including $S$-matrices. For level-two theories, we focus on the subexceptional series and found they are all supersymmetric and their fermionic characters satisfy a uniform fermionic MLDE. We also find many interesting coset constructions involving the Cvitanovi\'c--Deligne, sub and Severi exceptional series at level two.

Our analysis in this paper is primarily focused on the characters of these theories. It remains unclear, however, whether one can fully define the conformal field theory with the central charge given in magic triangle, or the correlation functions, for some of the theories. For example it would be interesting to find out whether the two atomic models $\mathrm{Vir}^{\rm eff}_{5,2}$ and $ \mathrm{Vir}^{\rm eff}_{5,3}$ are sensible conformal field theories with central charge $2/5$ and $3/5$ with proper Hilbert space and correlation functions.

We have not explored much of the dimension of irreducible representations for the magic triangle in this work. The magic coset we have uncovered suggest that the representations of all algebras appearing in the magic triangle are intricately interconnected. It would therefore be interesting to find some simple universal formulas for the dimensions  of these representations, besides the adjoint representation and some simpler representations appearing as those for non-vacuum primary operators.

Finally, although our discussion has focused on WZW models at positive integer levels, theories at fractional and negative levels are also of considerable interest. For instance, it is well known that the Cvitanovi\'c--Deligne exceptional series at level $-h^\vee/6-1$ gives rise to a celebrated family of quasi-lisse vertex operator algebras, whose vacuum characters satisfy a one-parameter family of second-order MLDEs \cite{Beem:2017ooy,AK}.
It is intriguing to consider whether this phenomenon can be  generalized to other exceptional series, or even to the entire magic triangle.

\section*{Acknowledgements}
We thank Chongying Dong, Heeyeon Kim, Ching Hung Lam, Jaewon Song, Haowu Wang, Xin Wang, Yinan Wang and Don Zagier for discussion. 
KL is supported in part by the BIMSA start-up fund and the Beijing Natural Science Foundation International Scientist Project No.IS25024. KS is supported by China NSF grant No.1250010384.

\appendix
\section{Explicit data of the extended magic triangle}\label{sec:A}
We collect the dual Coxeter number of dimension of each element of the extended magic triangle in Table \ref{tb:MTdata0}. We collect the dimensions of $R_1,R_2,R_3$ for all level-one theories  in Table \ref{tb:MTdataR}. 
The central charge, nonzero conformal weights and degeneracies for all level-one theories are summarized  in Table \ref{tb:MTdata}. Owing to  the symmetry exchanging  $\mu$ and $\nu$, we display the $\mu\le\nu$ portion of the triangle.

\begin{table}[ht]
\def\arraystretch{2}

	\centering
	\resizebox{\linewidth}{!}{
	\begin{tabular}{cccccccccc|c}
& & & &$ \{-1,0\} $&$ \left\{\frac{1}{4},3\right\} $& & & &  &$3/2$\\
 &  & &$ \{-1,0\} $&$ \{0,2\}$  & $\{\frac32,8\}$ & $ \{3,16\} $ & & & &2 \\
 & &$ \{0,1\} $&$ \left\{\frac{2}{3},3\right\} $&$ \{2,9\} $&$ \{4,21\} $&$ \{6,35\} $&$ \{10,66\} $& & & 3 \\
 &$ \left\{\frac{2}{3},1\right\} $&$ \left\{\frac{3}{2},4\right\} $&$ \left\{\frac{7}{3},8\right\} $&$ \{4,18\} $&$ \left\{\frac{13}{2},36\right\} $&$ \{9,56\} $&$ \{14,99\} $&$ \{19,144\} $& &4 \\
 $\left\{\frac{3}{2},1\right\} $&$ \{2,3\} $&$ \{3,8\} $&$ \{4,14\} $&$ \{6,28\} $&$ \{9,52\} $&$ \{12,78\} $&$ \{18,133\} $&$ \{24,190\} $&$ \{30,248\}$ & 5\\
\hline
 $1/4$ & $1/3$ & $1/2 $& $2/3$ & $1$ & $3/2$ & $2$ & $3$ & $4$ & $5$ & $\nu\backslash \mu$  \\  
 \end{tabular}}
        \caption{$\{h^\vee,\dim\}$ associated to the extended magic triangle.}
			\label{tb:MTdata0}
		\end{table}

\begin{table}[ht]
\def\arraystretch{2}

	\centering
	\resizebox{\linewidth}{!}{
	\begin{tabular}{c ccccccccc|c}
& & & & $ \left\{1,1,2\right\}$ & $ \left\{3,5,5\right\}$ & & & & &$3/2$ \\
 & & & $ \left\{1,1,1\right\}$ & $ \left\{1,2,3\right\}$ & $ \left\{3,8,6\right\}$ & $ \left\{ 3,9,9\right\}$ & & & & $2$\\
 & & $ \left\{1,1,2\right\}$ & $ \left\{2,3,4\right\}$ & $ \left\{2,4,8\right\}$ & $ \left\{6,14,14\right\}$ & $ \left\{ 6,15,20\right\}$ & $ \left\{12,32,32\right\}$ & & & $3$ \\
 & $ \left\{1,1,1\right\}$ & $ \left\{1,2,3\right\}$ & $ \left\{2,5,5\right\}$ & $ \left\{2,6,9\right\}$ & $ \left\{6,20,15\right\}$ & $ \left\{6,21,21\right\}$ & $ \left\{12,44,33\right\}$ & $ \left\{ 12,45,45\right\}$ &  & $4$ \\
 $ \left\{1\right\}$ & $ \left\{2\right\}$ & $ \left\{3\right\}$ & $ \left\{7\right\}$ & $ \left\{8\right\}$ & $ \left\{26\right\}$ & $ \left\{27\right\}$ & $ \left\{56\right\}$ & $ \left\{57\right\}$ & $ -$ & $5$ \\ \hline

 $1/4$ & $1/3$ & $1/2 $& $2/3$ & $1$ & $3/2$ & 2  & 3 & 4 & 5 & $\nu\backslash \mu$  \\  
 \end{tabular}}
        \caption{Dimension $\{R_1,R_2,R_3\}$ associated to the extended magic triangle at level $1$.}
			\label{tb:MTdataR}
		\end{table}

\begin{table}[ht]
\def\arraystretch{2}

	\centering
	\resizebox{\linewidth}{!}{
	\begin{tabular}{c ccccccccc|c}
& & & & $ \left\{\frac{6}{5},(\frac{1}{10})_3,(\frac{1}{2})_3,\frac{3}{5}\right\}$ & $ \left\{\frac{12}{5},\frac{1}{5},\frac{3}{5},\frac{3}{5}\right\}$ & & & & &$3/2$ \\
 & & & $ \!\!\!\left\{\frac{4}{5},(\frac{1}{15})_2,\frac{2}{5},(\frac{2}{3})_2\right\}\!\!$ & $ \left\{2,(\frac{1}{6})_6,(\frac{1}{2})_3,(\frac{2}{3})_2\right\}$ & $\! \left\{\frac{16}{5},(\frac{4}{15})_2,\frac{3}{5},(\frac{2}{3})_2\right\}\!$ & $\! \left\{4,(\frac{1}{3})_4,(\frac{2}{3})_2,(\frac{2}{3})_2\right\}\!$ & & & & $2$\\
 & & $ \left\{1,(\frac{1}{12})_2,(\frac{1}{3})_2,\frac{3}{4}\right\}$ & $ \left\{\frac{9}{5},\frac{3}{20},\frac{2}{5},\frac{3}{4}\right\}$ & $ \left\{3,(\frac{1}{4})_3,(\frac{1}{2})_3,\frac{3}{4}\right\}$ & $ \left\{\frac{21}{5},\frac{7}{20},\frac{3}{5},\frac{3}{4}\right\}$ & $ \left\{5,(\frac{5}{12})_2,(\frac{2}{3})_2,\frac{3}{4}\right\}$ & $ \left\{6,\frac{1}{2},\frac{3}{4},\frac{3}{4}\right\}$ & & & $3$ \\
 & $ \left\{\frac{3}{5},\frac{1}{20},\frac{1}{4},\frac{4}{5}\right\}$ & $ \left\{\frac{8}{5},(\frac{2}{15})_2,(\frac{1}{3})_2,\frac{4}{5}\right\}$ & $ \left\{\frac{12}{5},\frac{1}{5},\frac{2}{5},\frac{4}{5}\right\}$ & $ \left\{\frac{18}{5},(\frac{3}{10})_3,(\frac{1}{2})_3,\frac{4}{5}\right\}$ & $ \left\{\frac{24}{5},\frac{2}{5},\frac{3}{5},\frac{4}{5}\right\}$ & $ \left\{\frac{28}{5},(\frac{7}{15})_2,(\frac{2}{3})_2,\frac{4}{5}\right\}$ & $ \left\{\frac{33}{5},\frac{11}{20},\frac{3}{4},\frac{4}{5}\right\}$ & $ \left\{\frac{36}{5},\frac{3}{5},\frac{4}{5},\frac{4}{5}\right\}$ &  & $4$ \\
 $\!\!\!\! \left\{\frac{2}{5},\underline{\frac{1}{30}},\frac{1}{5},\underline{\frac{5}{6}}\right\}\!\!$ & $ \left\{1,\underline{\frac{1}{12}},\frac{1}{4},\underline{\frac{5}{6}}\right\}$ & $ \left\{2,\underline{\frac{1}{6}},(\frac{1}{3})_2,\underline{\frac{5}{6}}\right\}$ & $ \left\{\frac{14}{5},\underline{\frac{7}{30}},\frac{2}{5},\underline{\frac{5}{6}}\right\}$ & $ \left\{4,\underline{\frac{1}{3}},(\frac{1}{2})_3,\underline{\frac{5}{6}}\right\}$ & $ \left\{\frac{26}{5},\underline{\frac{13}{30}},\frac{3}{5},\underline{\frac{5}{6}}\right\}$ & $ \left\{6,\underline{\frac{1}{2}},(\frac{2}{3})_2,\underline{\frac{5}{6}}\right\}$ & $ \left\{7,\underline{\frac{7}{12}},\frac{3}{4},\underline{\frac{5}{6}}\right\}$ & $\! \left\{\frac{38}{5},\underline{\frac{19}{30}},\frac{4}{5},\underline{\frac{5}{6}}\right\}\!$ & $\! \left\{8,\underline{\frac{2}{3},\frac{5}{6},\frac{5}{6}}\right\}\!$ & $5$ \\ \hline

 $1/4$ & $1/3$ & $1/2 $& $2/3$ & $1$ & $3/2$ & 2  & 3 & 4 & 5 & $\nu\backslash \mu$  \\  
\end{tabular}}
        \caption{CFT data $\{c,h_1,h_2,h_3\}$ associated to the extended magic triangle at level $1$. Those with underlines indicate spurious solutions of the 4th order MLDE.}
			\label{tb:MTdata}
		\end{table}

\section{S-matrices}\label{sec:B0}
In this section, we collect the $S$-matrices for the two characters $\chi_0,  \chi_1$ of the Mathur--Mukhi--Sen series, namely  the extended Cvitanovi\'c--Deligne exceptional series at level one,  as reviewed in Section \ref{sec:MMS}.  
The $2\times 2$ $S$-matrix $S_{(\nu)} $ for the two characters of  $\cT^{(\nu,5)} $, which leaves the partition function \eqref{eq:CDpart} invariant,  satisfies $
    S_{(\nu)}= S_{(1/\nu)}. $
The $S$-matrices for 
  $\nu = 1/4,1/3,1/2,2/3,1$, corresponding respectively to the theories   ${\rm Vir}_{5,2}^{\rm eff}$, $(A_1)_1$, $(A_2)_1$, $(G_2)_1$, $(D_4)_1$,    are given by
\begin{align}
  & S_{(1/4)} =  \frac{\varphi^{1/2} }{5^{1/4} } \left(\begin{array}{cc}
        1 &  \varphi^{-1} \\
       \varphi^{-1}   & - 1
    \end{array}\right), \quad S_{(1/3)} =  \frac{1}{\sqrt{2}} \left(\begin{array}{cr}
      1  &  1 \\
        1  & -1
    \end{array}\right), \quad S_{(1/2)}  = \frac{1}{\sqrt{3}} \left(\begin{array}{cr}
      1  &  2 \\
       1 & -1
    \end{array}\right), \nonumber \\
    & S_{(2/3)}= \frac{1}{5^{1/4}\varphi^{1/2}} \left(\begin{array}{cc}
       1 & \varphi    \\
        \varphi & -1
    \end{array}\right), \quad S_{(1)}=\frac{1}{2} \left(\begin{array}{cr}
      1  &  3 \\
        1  & -1
    \end{array}\right),  
\end{align}
where $\varphi=(\sqrt{5}+1)/2$ is the golden ratio. The modular invariant partition function is $Z=|\chi_0|^2+ \cN|\chi_1|^2$ and $\cN$ is the degeneracy of the non-vacuum primary fields. 
Note that 
\begin{align}
    S_{(\nu)}^T \cdot\left(\begin{array}{cc}
    1 & 0 \\ 0 & \cN_{(\nu)}
    \end{array}\right)\cdot S_{(\nu)} = \left(\begin{array}{cc}
    1 & 0 \\ 0 & \cN_{(\nu)}. 
    \end{array}\right)
\end{align}
In addition for arbitrary $\nu$ we have $
    \det S_{(\nu)} =-1. $

For the  cases with degeneracy, namely $\nu=1/2,1,2$, we can have extended $S$-matrices, which are given by
\begin{align}
    S_{(1/2)}=\frac{1}{\sqrt{3}} \left(\begin{array}{ccc}
      1  &  1 & 1 \\
        1  & w & w^2 \\
        1 & w^2 & w 
    \end{array}\right)  , \quad  \ S_{(1)}=  \frac{1}{2} \left(\begin{array}{crrr}
      1  &  1 & 1 & 1 \\
        1  & 1 & -1 & -1 \\
        1 & -1 & 1 & -1 \\
        1 & -1 & -1 & 1
    \end{array}\right),   
\end{align}
where $w=e^{2\pi i /3}$.
Here all extended $S$-matrices are symmetric and unitary. From \eqref{eq:CDcoset}, the invariance of the pairs leads to the extended $S$-matrices of $\cT^{(\nu,5)}$ and $\cT^{(1/\nu,5)}$ are related by 
$ S_{(\nu)} = S_{(1/\nu)}^\dagger.$
The   $S$-matrix of   $\cT^{(\mu,\nu)}$   in full $(1+\cN_{(\mu)})\times (1+\cN_{(\nu)})$ form  is then 
\begin{align}
   \left(\begin{array}{cc}\chi_0 &   \chi_2 \\   \chi_3 &   \chi_1 \end{array}\right)^{(\mu,\nu)} \!\!\!\! (-1/\tau)\ = \ 
S_{(\mu)} \cdot \left(\begin{array}{cc}\chi_0 &   \chi_2 \\    \chi_3 &   \chi_1 \end{array}\right)^{(\mu,\nu)}  \!\!\!\!(\tau) \cdot S_{(\nu)}^\dagger.  
\end{align}

\section{Spurious solutions of MLDE}\label{sec:B}

For each $\mu$, in addition to the two genuine characters of $\mathcal{T}^{(5,\mu)}$, there exist two spurious solutions of the fourth order MLDE \eqref{MLDE1} whose q-expansion have coefficients with unbounded denominators. As an illustrative example, for $\mu=1/4$, the two genuine solutions are the Lee--Yang characters $\chi_0,\chi_{1/5}$, while the two spurious solutions are given by
\begin{align}\nonumber
 f_{\frac{1}{30}}   =&\   q^{\frac{1}{60}}\Big(1-\frac{2 q}{5}+\frac{q^2}{11}+\frac{26 q^3}{85}+\frac{434 q^4}{1265}+\frac{9824 q^5}{27115}+\dots\Big) , \\ \nonumber
 f_{\frac{5}{6}}   =& \  q^{\frac{49}{60}}\Big( 1+\frac{38 q}{33}+\frac{371 q^2}{561}+\frac{22558 q^3}{12903}+\frac{383219 q^4}{374187}+\dots\Big).
\end{align}
Although these two solutions are not themselves RCFT characters, we observe that they are closely related to the characters of the $(\Eh)_1$ theory via the relations
\begin{align}\nonumber
-180\Delta_{\frac14} f_{\frac{1}{30}}   =&\,  \eta^8 \chi_0^{(\Eh)_1}   , \\ \nonumber
108 \Delta_{\frac14}  f_{\frac{5}{6}}   =& \,  \eta^8 \chi_{\frac45}^{(\Eh)_1} .
\end{align}
Here $\Delta_\mu$ is the second order differential operator defined in \eqref{eq:deltanu}. These relations are straightforward to prove,  since $\Delta_4$ annihilates both sides of the equations. If we define $\chi_{\frac{1}{30}}=R_1 f_{\frac{1}{30}}$ and $\chi_{\frac{5}{6}}=R_3f_{\frac{5}{6}}$, and choose the prefactors   such that $R_3/R_1=3/5$, we obtain the  bilinear relation
\begin{equation}
   \chi_0\chi_{\frac{1}{30}}-\chi_{\frac{1}{5}}\chi_{\frac{5}{6}}=R_1  .
\end{equation}
similar to the magic bilinear relation \eqref{chiid}. This choice also ensures  the relation \eqref{eq:RRR} holds. For other $\mu\neq 5$,   similar relations can be derived.

The level one $(E_8)_1$ theory   has a single character $\chi_0=j^{1/3}(\tau)$ where $j(\tau)$ is the Klein $j$-invarant. 
The vacuum character satisfies the first order meromorphic MLDE,
\be   \Big( D + \frac{E_6}{3E_4}\Big)j^{1/3}=0. \ee
Accordinngly, the  second-order MLDE \eqref{KZ} for the $E_8$ theory factorizes  as
\begin{equation}
    D^2 -\frac{E_4}{6}= \Big( D-\frac{E_6}{3E_4}\Big)\Big(D+\frac{E_6}{3E_4}\Big). 
\end{equation}
For $\mu,\nu=5$, the fourth order MLDE \eqref{MLDE1} becomes
\begin{equation}
  \left[D^4-\frac{1}{6}E_4D^2+\frac19 E_6D-\frac{1}{36}E_4^2\right]\chi=\left[D^2(D^2-\frac16 E_4)\right]\chi=0  .
\end{equation}
Besides $\chi_0=j^{1/3}$ the vacuum character of $(E_8)_1$, this MLDE has two more spurious solutions
\begin{align}\nonumber
 f_{\frac{2}{3}}   =&\  q^{\frac{1}{3}}\Big(1+\frac{712 q}{25}+\frac{75436 q^2}{275}+\frac{95654016 q^3}{51425} +\dots\Big) , \\ \nonumber
 f_{\frac{5}{6}}   =&\ q^{\frac{1}{2}}\Big( 1+\frac{228 q}{11}+\frac{34938 q^2}{187}+\frac{5163352 q^3}{4301}+\dots\Big).
\end{align}
These solutions are not characters, but satisfy the following simple differential equations
\begin{align}\nonumber
-9 \Big(D^2-\frac16 E_4\Big)f_{\frac{2}{3}}   =&\  \eta^8  , \\ \nonumber
 \frac{6}{5}\Big(D+\frac{E_6}{3E_4}\Big)f_{\frac{5}{6}}   =&\ \eta^4 j^{-1/3} .
\end{align}
We find a bilinear relation for the three solutions
\begin{equation}
   \chi_0f_{\frac{2}{3}}-\frac{2^8\cdot 3^3}{5^2}f_{\frac{5}{6}}^2=1  ,
\end{equation}
similar to the magic bilinear relation \eqref{chiid}.


\end{document}